\documentclass[10pt,conference,letterpaper]{IEEEtran}
\usepackage{times}
\usepackage{algorithm}
\usepackage{algorithmic}
\usepackage{graphicx}
\usepackage{url}
\usepackage{epsfig}
\usepackage{amsmath,amssymb}
\usepackage{subfigure}
\usepackage{float}
\usepackage{balance}

\newtheorem{lemma}{Lemma}

\newtheorem{corollary}{Corollary}

\newtheorem{example}{Example}[section]

\newtheorem{defn}{Definition}[section]
\newtheorem{thm}{Theorem}[section]

\newtheorem{observation}{Observation}[section]
\newtheorem{proposition}{Proposition}[section]

\floatstyle{ruled}
\newfloat{procedure}{t}{lop}
\floatname{procedure}{Procedure}
\newfloat{algo}{t}{lop}
\floatname{algo}{Algorithm}

\newcommand{\blob}{\rule[.2ex]{1ex}{1ex}}
\newcommand{\bc}{\begin{center}}
\newcommand{\ec}{\end{center}}
\newcommand{\ra}{\rightarrow}

\newcommand{\bi}{\begin{itemize}}
\newcommand{\ei}{\end{itemize}}
\newcommand{\be}{\begin{enumerate}}
\newcommand{\ee}{\end{enumerate}}
\newcommand{\bd}{\begin{description}}
\newcommand{\ed}{\end{description}}
\newcommand{\bo}{\begin{observation}}
\newcommand{\eo}{\end{observation}}
\newcommand{\bt}{\begin{thm}}
\newcommand{\et}{\end{thm}}
\newcommand{\bl}{\begin{lemma}}
\newcommand{\el}{\end{lemma}}
\newcommand{\bp}{\begin{proposition}}
\newcommand{\ep}{\end{proposition}}
\newcommand{\bprop}{\begin{property}}
\newcommand{\eprop}{\end{property}}
\newcommand{\bde}{\begin{defn}}
\newcommand{\ede}{\end{defn}}

\newcommand{\bx}{\begin{example}}
\newcommand{\ex}{\end{example}}

\newcommand{\bco}{\begin{corollary}}
\newcommand{\eco}{\end{corollary}}

\newcommand{\btab}{\begin{tabbing}}
\newcommand{\etab}{\end{tabbing}}

\newcommand{\squishenu}{
   \begin{enumerate}
    { \setlength{\itemsep}{0pt}      \setlength{\parsep}{3pt}
      \setlength{\topsep}{3pt}       \setlength{\partopsep}{0pt}
      \setlength{\leftmargin}{1.5em} \setlength{\labelwidth}{1em}
      \setlength{\labelsep}{0.5em} } }
\newcommand{\squishenuend}{
    \end{enumerate}  }
\newcommand{\squishlist}{
   \begin{list}{$\bullet$}
    { \setlength{\itemsep}{0pt}      \setlength{\parsep}{3pt}
      \setlength{\topsep}{3pt}       \setlength{\partopsep}{0pt}
      \setlength{\leftmargin}{1.5em} \setlength{\labelwidth}{1em}
      \setlength{\labelsep}{0.5em} } }
\newcommand{\squishlisttwo}{
   \begin{list}{$\bullet$}
    { \setlength{\itemsep}{0pt}    \setlength{\parsep}{0pt}
      \setlength{\topsep}{0pt}     \setlength{\partopsep}{0pt}
      \setlength{\leftmargin}{2em} \setlength{\labelwidth}{1.5em}
      \setlength{\labelsep}{0.5em} } }
\newcommand{\squishlistend}{
    \end{list}  }

\title{Differentially Private Trajectory Data Publication}
\author{%
{Rui Chen{\small $~^{\#1}$}, Benjamin C. M. Fung{\small $~^{*2}$}, Bipin C. Desai{\small $~^{\#3}$} }%
\vspace{1.6mm}\\
\fontsize{10}{10}\selectfont\itshape
$^{\#}$\,Department of Computer Science and Software Engineering, Concordia University\\
Montreal, Quebec, Canada\\
\fontsize{9}{9}\selectfont\ttfamily\upshape
$^{1}$\,ru\_che@encs.concordia.ca\\
$^{3}$\,bcdesai@cs.concordia.ca%
\vspace{1.2mm}\\
\fontsize{10}{10}\selectfont\rmfamily\itshape
$^{*}$\,Concordia Institute for Information Systems Engineering, Concordia University\\
Montreal, Quebec, Canada\\
\fontsize{9}{9}\selectfont\ttfamily\upshape
$^{2}$\,fung@ciise.concordia.ca
}
\begin{document}
\maketitle
\begin{abstract}
With the increasing prevalence of location-aware devices, trajectory data has been generated and collected in various application domains. Trajectory data carries rich information that is useful for many data analysis tasks. Yet, improper publishing and use of trajectory data could jeopardize individual privacy. However, it has been shown that existing privacy-preserving trajectory data publishing methods derived from \emph{partition-based} privacy models, for example $k$-anonymity, are unable to provide sufficient privacy protection. 

In this paper, motivated by the data publishing scenario at the \emph{Soci\'{e}t\'{e} de transport de Montr\'{e}al} (STM), the public transit agency in Montreal area, we study the problem of publishing trajectory data under the rigorous \emph{differential privacy} model. We propose an efficient \emph{data-dependent} yet differentially private sanitization algorithm, which is applicable to different types of trajectory data. The efficiency of our approach comes from adaptively narrowing down the output domain by building a noisy prefix tree based on the underlying data. Moreover, as a post-processing step, we make use of the inherent constraints of a prefix tree to conduct constrained inferences, which lead to better utility. This is the first paper to introduce a practical solution for publishing large volume of trajectory data under differential privacy. We examine the utility of sanitized data in terms of count queries and frequent sequential pattern mining. Extensive experiments on real-life trajectory data from the STM demonstrate that our approach maintains high utility and is scalable to large trajectory datasets.
\end{abstract}

%
\section{Introduction}\label{sec:Introduction}
Over the last few years, location-aware devices, such as RFID tags, cell phones, GPS navigation systems, and point of sale (POS) terminals, have been widely deployed in various application domains. Such devices generate large volume of trajectory data that could be used for many important data analysis tasks, such as marketing analysis~\cite{UAW06}, long-term network planning~\cite{PK08}, customer behavior analysis~\cite{BCM06}, and demand forecasting~\cite{UAW06}. However, trajectory data often contains sensitive personal information, and improper publishing and use of trajectory data may violate individual privacy. The privacy concern of publishing trajectory data is best exemplified by the case of the \emph{Soci\'{e}t\'{e} de transport de Montr\'{e}al} (STM, http://www.stm.info), the public transit agency in Montreal area.

In 2007, the STM deployed the smart card automated fare collection (SCAFC) system as a secure method of user validation and fare collection. In addition to revenue collection, the system generates and collects passengers' trajectory data every day. Transit information, such as smart card ID and station ID, is collected when a passenger swipes his smart card at a SCAFC system, and is then stored in a central database management system, where the transit information of a passenger is organized as an ordered list of $stations$, a kind of trajectory data (see a formal definition in Section~\ref{subsec:trajectory database}). Periodically, the IT department of the STM shares such trajectory data with other departments, e.g., the marketing department, for basic data analysis, and publishes its trajectory data to external research institutions for more complex data mining tasks. According to the preliminary research~\cite{BCM06}, \cite{MMC09}, \cite{CMB07}, the STM can substantially benefit from trajectory data analysis at strategic, tactical, and operational levels. Yet, it has also realized that the nature of trajectory data is raising major privacy concerns on the part of card users in information sharing~\cite{MMC09}. This fact has been an obstacle to further conduct trajectory data analysis tasks and even perform regular commercial operations. Similarly, many other sectors, for example cell phone communication and credit card payment~\cite{RC01}, have been facing the dilemma in trajectory data publishing and individual privacy protection.

The privacy concern in trajectory data sharing has spawned some research~\cite{ABN08}, \cite{TM08}, \cite{YBLW09}, \cite{HXODN10}, \cite{CFMD11}, \cite{MA10} on privacy-preserving trajectory data publishing based on \emph{partition-based} privacy models~\cite{GKS08}, for example $k$-anonymity~\cite{Sweeney02} (or $(k, \delta)$-anonymity~\cite{ABN08}) and confidence bounding~\cite{TM08}, \cite{CFMD11}. However, many types of privacy attacks, such as \emph{composition attack}~\cite{GKS08}, \emph{deFinetti attack}~\cite{Kifer09} and \emph{foreground knowledge attack}~\cite{WFWYP}, have been identified on the approaches derived using the partition-based privacy models, demonstrating their vulnerability to an adversary's background knowledge. Due to the deterministic nature of partition-based privacy models, it is foreseeable that more types of privacy attacks could be discovered on these privacy models in the future. Consequently, in recent years \emph{differential privacy}~\cite{DMNS06} has become the de facto successor to partition-based privacy models. Differential privacy provides provable privacy guarantees independent of an adversary's background knowledge and computational power (this claim may not be valid in some special cases~\cite{KM11}, but is still correct for our scenario, as discussed in Section~\ref{subsec:analysis}). Differential privacy requires that any computation based on the underlying database should be insensitive to the change of a single record. Therefore, a record owner can be ensured that any privacy breach would not be a result of participating in a database.

The traditional \emph{non-interactive} approaches~\cite{BLR08}, \cite{DNRRV09}, \cite{XWG10}, \cite{XXY10} for generating differentially-private releases are \emph{data-independent} in the sense that all possible entries in the output domain need to be \emph{explicitly} considered no matter what the underlying database is. For high-dimensional data, such as trajectory data, this is computationally infeasible. Consider a trajectory database $\mathcal{D}$ with all locations drawn from a universe of size $m$. Suppose the maximum length of trajectories (the number of locations in a trajectory) in $\mathcal{D}$ is $l$. These approaches need to generate $\sum_{i=1}^{l}m^i = \frac{m^{l+1}-m}{m-1}$ output entries. For a trajectory database with $m=1,000$ and $l=20$, it requires to generate $10^{60}$ entries. Hence, these approaches are not computationally applicable with today's systems to real-life trajectory databases. 

Two very recent papers~\cite{MCFY11}, \cite{CMFDX11} point out that more efficient and more effective solutions could be achieved by carefully making use of the underlying database. We call such solutions \emph{data-dependent}. The general idea of data-dependent solutions is to adaptively narrow down the output domain by using \emph{noisy} answers obtained from the underlying database. However, the methods in~\cite{MCFY11}, \cite{CMFDX11} cannot be applied to trajectory data for two reasons. First, the methods in~\cite{MCFY11}, \cite{CMFDX11} require taxonomy trees to guide the data publication process. For trajectory data, there does not exist a logical taxonomy tree due to the \emph{sequentiality} among locations. Second, the methods only work for \emph{sets}, yet a trajectory may contain a \emph{bag} of locations. Therefore, non-trivial efforts are needed to develop a differentially private data publishing approach for trajectory data.

Protecting individual privacy is one aspect of sanitizing trajectory data. Another equally important aspect is preserving utility in sanitized data for data analysis. Motivated by the STM case, in this paper, we are particularly interested in two kinds of data mining tasks, namely \emph{count queries} (see a formal definition in Section~\ref{subsec:utility}) and \emph{frequent sequential pattern mining}~\cite{AS95}. Count queries, as a general data analysis task, are the building block of many more advanced data mining tasks. In the STM scenario, with accurate answers to count queries over sanitized data, data recipients can obtain the answers to questions, such as ``how many passengers have visited both stations \emph{Guy-Concordia} and \emph{McGill}~\footnote{\emph{Guy-Concordia} and \emph{McGill} are two metro stations on the green line of the Montreal metro network.} within the last week". Frequent sequential pattern mining, as a concrete data mining task, helps, for example, the STM better understand passengers' transit patterns and consequently allows the STM to adjust its network geometry and schedules in order to better utilize its existing resources. These utility requirements naturally demand a solution that publishes data, not merely data mining results.

\vspace{2mm}
\noindent\textbf{Contribution.} In this paper, we study the problem of publishing trajectory data that simultaneously protects individual privacy under the framework of differential privacy and provides high utility for different data mining tasks. This is the first paper that introduces a practical solution for publishing large volume of real-life trajectory data via differential privacy. The previous works~\cite{ABN08}, \cite{TM08}, \cite{YBLW09}, \cite{HXODN10}, \cite{CFMD11}, \cite{MA10} on privacy-preserving trajectory data publishing cannot be used to achieve differential privacy because of their deterministic nature. We summarize the major contributions of the paper as follows.

\begin{itemize}
\item	We propose a \emph{non-interactive data-dependent} sanitization algorithm of runtime complexity $O(|\mathcal{D}|\cdot |\mathcal{L}|)$ to generate a differentially private release for trajectory data, where $|\mathcal{D}|$ is the size of the underlying database $\mathcal{D}$ and $|\mathcal{L}|$ is the size of the location universe. The efficiency is achieved by constructing a \emph{noisy} prefix tree, which adaptively guides the algorithm to circumvent certain output sub-domains based on the underlying database.
\item We design a statistical process for efficiently constructing a noisy prefix tree under Laplace mechanism. This is vital to the scalability of processing datasets with large location universe sizes.
\item	We make use of two sets of inherent constraints of a prefix tree to conduct constrained inferences, which helps generate a more accurate release. This is the first paper of applying constrained inferences to \emph{non-interactive} data publishing.
\item	We conduct an extensive experimental study over the real-life trajectory dataset from the STM. We examine utility of sanitized data for two different data mining tasks, namely count queries (a generic data analysis task) and frequent sequential pattern mining (a concrete data mining task). We demonstrate that our approach maintains high utility and is scalable to large volume of real-life trajectory data.
\end{itemize}

The rest of the paper is organized as follows. Section~\ref{sec:related work} reviews related work. Section~\ref{sec:preliminaries} provides the preliminaries for our solution. A two-stage sanitization algorithm for trajectory data is proposed in Section~\ref{sec:sanitization}, and comprehensive experimental results are presented in Section~\ref{sec:experiment}. Finally, we conclude the paper in Section~\ref{sec:conclusion}.


\section{Related Work} \label{sec:related work}

In this section, we review the state of the art of privacy-preserving trajectory data publishing techniques and recent applications of differential privacy.

\subsection{Privacy-Preserving Trajectory Data Publishing} \label{subsec:privacy preserving trajectory data publishing}

Due to the ubiquitousness of trajectory data, some recent works~\cite{ABN08}, \cite{TM08}, \cite{YBLW09}, \cite{HXODN10}, \cite{CFMD11}, \cite{MA10} have started to study privacy-preserving trajectory data publishing from different perspectives. Abul et al.~\cite{ABN08} propose the $(k, \delta)$-anonymity model based on the inherent imprecision of sampling and positioning systems, where $\delta$ represents the possible location imprecision. The general idea of~\cite{ABN08} is to modify trajectories by \emph{space translation} so that $k$ different trajectories co-exist in a cylinder of the radius $\delta$. Terrovitis and Mamoulis~\cite{TM08} model an adversary's background knowledge as a set of projections of trajectories in a trajectory database, and consequently propose a data suppression technique that limits the confidence of inferring the presence of a location in a trajectory to a pre-defined probability threshold. Yarovoy et al.~\cite{YBLW09} propose to $k$-anonymize a moving object database (MOD) by considering timestamps as the quasi-identifiers (QIDs). Adversaries are assumed to launch privacy attacks based on \emph{attack graphs}. Their approach first identifies anonymization groups and then generalizes the groups to common regions according to the QIDs while achieving minimal information loss. Monreale et al.~\cite{MA10} present an approach based on spatial generalization in order to achieve $k$-anonymity. The novelty of their approach lies in a generalization scheme that depends on the underlying trajectory dataset rather than a fixed grid hierarchy.

Hu et al.~\cite{HXODN10} present the problem of $k$-anonymizing a trajectory database with respect to a sensitive event database. The goal is to make sure that every event is shared by at least $k$ users. Specifically, they develop a new generalization mechanism known as \emph{local enlargement}, which achieves better utility than conventional hierarchy- or partition-based generalization. Chen et al.~\cite{CFMD11} consider the emerging trajectory data publishing scenario, in which users' sensitive attributes are published with trajectory data and consequently propose the $(K, C)_L$-privacy model that thwarts both identity linkages on trajectory data and attribute linkages via trajectory data. They develop a generic solution for various data utility metrics by use of \emph{local suppression}. All these approaches~\cite{ABN08}, \cite{TM08}, \cite{YBLW09}, \cite{HXODN10}, \cite{CFMD11}, \cite{MA10} are built based on partition-based privacy models, and therefore are not able to provide sufficient privacy protection for trajectory data. The major contribution of our paper is the use of differential privacy, which provides significantly stronger privacy guarantees.

\subsection{Applications of Differential Privacy}

In the last few years, differential privacy has been employed in various applications. Currently most of the research on differential privacy concentrates on the \emph{interactive setting} with the goal of either reducing the magnitude of added noise~\cite{HRMS10}, \cite{LHRMM10}, \cite{RR10}, \cite{XBHG11} or releasing certain data mining results ~\cite{BCDKMT07, BLST2010, FS10, KKMN09, MKAGV08}. Dwork~\cite{Dwork2011} provides an overview of recent works on differential privacy. 

The works closest to ours are by Blum et al.~\cite{BLR08}, Dwork et al.~\cite{DNRRV09}, Xiao et al.~\cite{XWG10}, Xiao et al.~\cite{XXY10}, Mohammed et al.~\cite{MCFY11}, and Chen et al.~\cite{CMFDX11}. All these works consider \emph{non-interactive} data publishing under differential privacy. Blum et al.~\cite{BLR08} demonstrate that it is possible to release \emph{synthetic} private databases that are useful for all queries over a discretized domain from a concept class with polynomial Vapnik-Chervonenkis dimension~\footnote{Vapnik-Chervonenkis dimension is a measure of the complexity of a concept in the class.}. However, their mechanism is not efficient, taking runtime complexity of $superpoly(|\mathcal{C}|, |I|)$, where $|\mathcal{C}|$ is the size of a concept class and $|I|$ the size of the universe. Dwork et al. ~\cite{DNRRV09} propose a recursive algorithm of generating a \emph{synthetic} database with runtime complexity of $poly(|\mathcal{C}|, |I|)$. This improvement, however, is still insufficient to handle real-life trajectory datasets due to the exponential size of $|\mathcal{C}|$. Xiao et al.~\cite{XWG10} propose a wavelet-transformation based approach for \emph{relational data} to lower the magnitude of noise, rather than adding independent Laplace noise. Xiao et al.~\cite{XXY10} propose a two-step algorithm for \emph{relational data}. It first issues queries for \emph{every} possible combination of attribute values to the PINQ interface~\cite{McSherry09}, and then produces a generalized output based on the perturbed results. Similarly, the algorithms~\cite{XWG10}, \cite{XXY10} need to process all possible entries in the entire output domain, giving rise to a scalability problem in the context of trajectory data.

Two very recent papers~\cite{MCFY11}, \cite{CMFDX11} point out that data-dependent approaches are more efficient and more effective for generating a differentially private release. Mohammed et al.~\cite{MCFY11} propose a generalization-based sanitization algorithm for \emph{relational data} with the goal of classification analysis. Chen et al.~\cite{CMFDX11} propose a probabilistic top-down partitioning algorithm for \emph{set-valued data}. Both approaches~\cite{MCFY11}, \cite{CMFDX11} make use of taxonomy trees to adaptively narrow down the output domain. However, due to the reasons mentioned in Section~\ref{sec:Introduction}, they cannot be applied to trajectory data, in which \emph{sequentiality} is a major concern. 


\section{Preliminaries}\label{sec:preliminaries}
In this section, we define a trajectory database and a prefix tree, review differential privacy, and present the utility requirements. The notational conventions are summarized in Table~\ref{table:notation}.

\begin{table}[t]
    \centering
    \caption{Notational conventions} \label{table:notation}
    \begin{tabular}{|l|l|}\hline
        \textbf{Symbol} & \textbf{Description} \\ \hline
        $\mathcal{A}$ & A privacy mechanism \\
        $\Delta f $ & The global sensitivity of the function $f$ \\
        $\epsilon, \bar{\epsilon}$ & The total privacy budget, a portion of privacy budget\\
        $\mathcal{L}, L_i$ & The location universe, a location in the universe\\
        $T, t_i$ & A trajectory, a location in a trajectory\\
        $ls(T)$ & The set of locations in trajectory $T$ \\
        $\mathcal{D}, \widetilde{\mathcal{D}}$ & A trajectory database, a sanitized database of $\mathcal{D}$\\
        $Q(\mathcal{D})$ & A count query over the database $\mathcal{D}$\\       
        $\mathcal{PT}$ & A prefix tree\\
        $Root(\mathcal{PT})$ & The virtual root of the prefix tree $\mathcal{PT}$\\
        $prefix(v, \mathcal{PT})$ & The prefix represented by the node $v$ of $\mathcal{PT}$\\
        $tr(v)$ & The set of trajectories with the prefix $prefix(v, \mathcal{PT})$\\
        $c(v), \widetilde{c}(v), \bar{c}(v)$ & The noisy count, intermediate estimate and consistent\\
        & estimate of $|tr(v)|$ respectively\\
        $s$ & Sanity bound\\
        $\mathcal{F}_k(\mathcal{D})$ & The top-$k$ most frequent sequential patterns of $\mathcal{D}$\\
        $|\mathcal{L}|, |T|, |\mathcal{D}|$ & The size of the location universe, a trajectory, and\\
        & a trajectory database respectively\\
        $k$ & The number of empty nodes that pass the boolean tests\\
        \hline
    \end{tabular}
\end{table}

\subsection{Trajectory Database}\label{subsec:trajectory database}

Let $\mathcal{L}=\{L_1, L_2, \cdots, L_{|\mathcal{L}|}\}$ be the universe of locations, where $|\mathcal{L}|$ is the size of the universe. Without loss of generality, we consider locations as discrete spatial areas in a map. For example, in the STM case, $\mathcal{L}$ represents all stations in the STM transportation network. This assumption also applies to many other types of trajectory data, e.g., purchase records, where a location is a store's address. We model a \emph{trajectory} as an ordered list of locations drawn from the universe. 

\vspace{2mm}
\bde[Trajectory] \em A trajectory $T$ of length $|T|$ is an ordered list of locations $T = t_1 \ra t_2 \ra \cdots \ra t_{|T|}$, where $\forall 1\leq i \leq |T|$, $t_i \in \mathcal{L}$.~\blob \em
\ede
\vspace{2mm}

A location may occur multiple times in $T$, and may occur consecutively in $T$. Therefore, given $\mathcal{L} = \{L_1, L_2, L_3, L_4\}$, $T = L_1 \ra L_2 \ra L_2$ is a valid trajectory. In some cases, a trajectory may include timestamps. We point out that our approach also works for this type of trajectory data and discuss the details in Section~\ref{subsec:extensibility}. A trajectory database is composed of a multiset of trajectories; each trajectory represents the movement history of a record owner. A formal definition is as follow.

\vspace{2mm}
\bde[Trajectory Database] \em A trajectory database $\mathcal{D}$ of size $|\mathcal{D}|$ is a multiset of trajectories $\mathcal{D} = \{D_1, D_2, \cdots, D_{|\mathcal{D}|}\}$.~\blob \em
\ede
\vspace{2mm}

Table~\ref{table:rawdata} presents a sample trajectory database with $\mathcal{L} = \{L_1, L_2, L_3, L_4\}$. In the rest of the paper, we use the term database and dataset interchangeably.

\begin{table}[t]
    \centering
    \caption{Sample trajectory database} \label{table:rawdata}
    \begin{tabular}{|l|l|}\hline
        \textbf{Rec. \#} & \textbf{Path} \\ \hline
        1 & $ L_1 \ra L_2 \ra L_3 $\\
        2 & $ L_1 \ra L_2 $\\
        3 & $ L_3 \ra L_2 \ra L_1 $ \\
        4 & $ L_1 \ra L_2 \ra L_4 $\\
        5 & $ L_1 \ra L_2 \ra L_3 $\\
        6 & $ L_3 \ra L_2 $\\
        7 & $ L_1 \ra L_2 \ra L_4 \ra L_1 $\\
        8 & $ L_3 \ra L_1 $\\
        \hline
    \end{tabular}
\end{table}

\subsection{Prefix Tree}
A trajectory database can be represented in a more compact way in terms of a \emph{prefix tree}. A prefix tree groups trajectories with the same prefix into the same branch. We first define a prefix of a trajectory below. 

\vspace{2mm}
\bde[Trajectory Prefix] \em A trajectory $S = s_1 \ra s_2 \ra \cdots \ra s_{|S|}$ is a prefix of a trajectory $T = t_1 \ra t_2 \ra \cdots \ra t_{|T|}$, denoted by $S \preceq T$, if and only if $|S|\leq |T|$ and $\forall 1 \leq i \leq |S|$, $s_i = t_i$.~\blob
\ede
\vspace{2mm}

For example, $L_1 \ra L_2$ is a prefix of $L_1 \ra L_2 \ra L_4 \ra L_3$, but $L_1 \ra L_4$ is \emph{not}. Note that a trajectory prefix is a trajectory per se. Next, we formally define a prefix tree below.

\vspace{2mm}
\bde[Prefix Tree] \em A prefix tree $\mathcal{PT}$ of a trajectory database $\mathcal{D}$ is a triplet $\mathcal{PT} = (V, E, Root(\mathcal{PT}))$, where $V$ is the set of nodes labeled with locations, each corresponding to a unique trajectory prefix in $\mathcal{D}$; $E$ is the set of edges, representing transitions between nodes; $Root(\mathcal{PT}) \in V$ is the virtual root of $\mathcal{PT}$. The unique trajectory prefix represented by a node $v \in V$, denoted by $prefix(v, \mathcal{PT})$, is an ordered list of locations starting from $Root(\mathcal{PT})$ to $v$.~\blob
\ede
\vspace{2mm}

\begin{figure}[t]
\centering
\scalebox{0.6}{
\includegraphics[bb= 105 58 371 268]{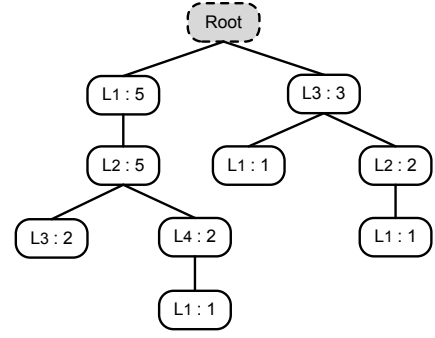}
}
\caption{The prefix tree of the sample data}\label{figure:prefixtree}
\end{figure}

Each node $v \in V$ of $\mathcal{PT}$ keeps a doublet in the form of $\langle tr(v), c(v)\rangle$, where $tr(v)$ is the set of trajectories in $\mathcal{D}$ having the prefix $prefix(v, \mathcal{PT})$, that is, $\{D \in \mathcal{D} : prefix(v, \mathcal{PT}) \preceq D\}$, and $c(v)$ is a noisy version of $|tr(v)|$ (e.g., $|tr(v)|$ plus Laplace noise). $tr(Root(\mathcal{PT}))$ contains all trajectories in $\mathcal{D}$. We call the set of all nodes of $\mathcal{PT}$ at a given depth $i$ a \emph{level} of $\mathcal{PT}$, denoted by $level(i, \mathcal{PT})$. $Root(\mathcal{PT})$ is at depth zero. Figure~\ref{figure:prefixtree} illustrates the prefix tree of the sample database in Table~\ref{table:rawdata}, where each node $v$ is labeled with its location and $|tr(v)|$.

\subsection{Differential Privacy}

Differential privacy is a relatively new privacy model stemming from the field of \emph{statistical disclosure control}. Differential privacy, in general, requires that the removal or addition of a single database record does not significantly affect the outcome of any analysis based on the database. Therefore, for a record owner, any privacy breach will not be a result of participating in the database since anything that can be learned from the database with his record can also be learned from the one without his record. We formally define differential privacy in the \emph{non-interactive} setting~\cite{BLR08} as follow.

\vspace{2mm}
\bde[$\epsilon$-differential privacy] \em A non-interactive privacy mechanism $\mathcal{A}$ gives $\epsilon$-differential privacy if for any database $\mathcal{D}_1$ and $\mathcal{D}_2$ differing on at most one record, and for any possible sanitized database $\widetilde{\mathcal{D}} \in Range(\mathcal{A})$,
\begin{align}
 Pr[\mathcal{A}(\mathcal{D}_1)= \widetilde{\mathcal{D}}]\leq e^{\epsilon}\times Pr[\mathcal{A}(\mathcal{D}_2)= \widetilde{\mathcal{D}}]
\end{align}
\noindent where the probability is taken over the randomness of $\mathcal{A}$.~\blob
\ede
\vspace{2mm}

Two principal techniques for achieving differential privacy have appeared in the literature, namely \emph{Laplace mechanism}~\cite{DMNS06} and \emph{exponential mechanism}~\cite{MT07}. A fundamental concept of both techniques is the \emph{global sensitivity} of a function~\cite{DMNS06} mapping underlying databases to (vectors of) reals.

\vspace{2mm}
\bde[Global Sensitivity] \em
For any function $f:\mathcal{D} \rightarrow \mathbb{R}^ d$, the sensitivity of $f$ is
\begin{align}
\Delta f & = \max_{\mathcal{D}_1, \mathcal{D}_2} || f(\mathcal{D}_1)-f(\mathcal{D}_2) ||_1
\end{align}
\noindent for all $\mathcal{D}_1, \mathcal{D}_2$ differing in at most one record.~\blob
\ede
\vspace{2mm}

Functions with lower sensitivity are more tolerant towards changes of a database and, therefore, allow more accurate differentially private mechanisms.

\vspace{2mm}
\noindent \textbf{Laplace Mechanism.} For the analysis whose outputs are real, a standard mechanism to achieve differential privacy is to add Laplace noise to the true output of a function. Dwork et al.~\cite{DMNS06} propose the Laplace mechanism which takes as inputs a database $\mathcal{D}$, a function $f$, and the privacy parameter $\epsilon$. The noise is generated according to a Laplace distribution with the probability density function (pdf) $p(x|\lambda)= \frac{1}{2\lambda}e^{-|x|/\lambda}$, where $\lambda$ is determined by both $\Delta f$ and the desired privacy parameter $\epsilon$.

\vspace{2mm}
\bt \em For any function $f:\mathcal{D} \rightarrow \mathbb{R}^ d$,
the mechanism $\mathcal{A}$
\begin{align}
\mathcal{A}(\mathcal{D}) = f(\mathcal{D}) + Laplace(\Delta f/\epsilon)
\end{align}
\noindent gives $\epsilon$-differential privacy.~\blob
\et
\vspace{2mm}

For example, for a single count query $Q$ over a dataset $\mathcal{D}$, returning $Q(\mathcal{D})+ Laplace(1/\epsilon)$ maintains $\epsilon$-differential privacy because a count query has a sensitivity 1.

\vspace{2mm}
\noindent \textbf{Exponential Mechanism.} For the analysis whose outputs are not real or make no sense after adding noise, McSherry and Talwar~\cite{MT07} propose the exponential mechanism that selects an output from the output domain, $r \in \mathcal{R}$, by taking into consideration its score of a given utility function $q$ in a differentially private manner. The exponential mechanism assigns exponentially greater probabilities of being selected to outputs of higher scores so that the final output would be close to the optimum with respect to $q$. The chosen utility function $q$ should be insensitive to changes of any particular record, that is, has a low sensitivity. Let the sensitivity of $q$ be $\Delta q= \max_{\forall r, \mathcal{D}_1, \mathcal{D}_2}$ $|q(\mathcal{D}_1,r)-q(\mathcal{D}_2,r)|$.

\vspace{2mm}
\bt \em Given a utility function $q: (\mathcal{D} \times \mathcal{R})
\rightarrow \mathbb{R}$ for a database $\mathcal{D}$, the mechanism $\mathcal{A}$,
\begin{align}
\mathcal{A}(\mathcal{D}, q) = \left\{return\ r\ with\ probability \propto exp({\frac{\epsilon q(\mathcal{D}, r)}{2\Delta q}})\right\}
\end{align}
\noindent gives $\epsilon$-differential privacy.~\blob
\et
\vspace{2mm}

\vspace{2mm}
\noindent \textbf{Composition Property.} For a sequence of computations, its privacy guarantee is provided by the composition properties. Any sequence of computations that each provides differential privacy in isolation also provides differential privacy in sequence, which is known as \emph{sequential composition}~\cite{McSherry09}.

\vspace{2mm}
\bt \label{thm:scomposition} \em Let $\mathcal{A}_i$ each provide $\epsilon_i$-differential privacy. A sequence of $\mathcal{A}_i(\mathcal{D})$ over the database $\mathcal{D}$ provides ($\sum_{i}\epsilon_i$)-differential privacy.~\blob
\et
\vspace{2mm}

In some special cases, in which a sequence of computations is conducted on \emph{disjoint} databases, the privacy cost does not accumulate, but depends only on the worst guarantee of all computations. This is known as \emph{parallel composition}~\cite{McSherry09}. This property could and should be used to obtain good performance.

\vspace{2mm}
\bt \label{thm:pcomposition} \em Let $\mathcal{A}_i$ each provide $\epsilon_i$-differential privacy. A sequence of $\mathcal{A}_i(\mathcal{D}_i)$ over a set of disjoint datasets $\mathcal{D}_i$ provides ($max(\epsilon_i)$)-differential privacy.~\blob
\et
\vspace{2mm}

\subsection{Utility Requirements}\label{subsec:utility}

The sanitized data is mainly used to perform two different data mining tasks, namely \emph{count query} and \emph{frequent sequential pattern mining}~\cite{AS95}. Count queries, as a general data analysis task, are the building block of many data mining tasks. We formally define count queries over a trajectory database below.

\vspace{2mm}
\bde[Count Query] \em
For a given set of locations $\mathbb{L}$ drawn from the universe $\mathcal{L}$, a count query $Q$ over a database $\mathcal{D}$ is defined to be $Q(\mathcal{D})=|\{D \in \mathcal{D}: \mathbb{L} \subseteq ls(D)\}|$, where $ls(D)$ returns the set of locations in $D$.~\blob
\ede
\vspace{2mm}

Note that sequentiality among locations is not considered in count queries, because the major users of count queries are, for example, the personnel of the marketing department of the STM, who are merely interested in users' presence in certain stations for marketing analysis, but \emph{not} the sequentiality of visiting. Instead, the preservation of sequentiality in sanitized data is examined by frequent sequential pattern mining. We measure the utility of a count query over the sanitized database $\widetilde{\mathcal{D}}$ by its \emph{relative error}~\cite{XWG10}, \cite{XBHG11}, \cite{CMFDX11} with respect to the true result over the original database $\mathcal{D}$, which is computed as: $$\frac{|Q(\widetilde{\mathcal{D}})-Q(\mathcal{D})|}{max\{Q(\mathcal{D}), s\}},$$ where $s$ is a \emph{sanity bound} used to mitigate the influences of the queries with extremely small \emph{selectivities}~\cite{XWG10}, \cite{XBHG11}, \cite{CMFDX11}.

For frequent sequential pattern mining, we measure the utility of sanitized data in terms of \emph{true positive}, \emph{false positive} and \emph{false drop}~\cite{ESAG04}. Given a positive number $k$, we denote the set of top $k$ most frequent sequential patterns identified on the original database $\mathcal{D}$ by $\mathcal{F}_k(\mathcal{D})$ and the set of frequent sequential patterns on the sanitized database $\widetilde{\mathcal{D}}$ by $\mathcal{F}_k(\widetilde{\mathcal{D}})$. True positive is the number of frequent sequential patterns in $\mathcal{F}_k(\mathcal{D})$ that are correctly identified in $\mathcal{F}_k(\widetilde{\mathcal{D}})$, that is, $|\mathcal{F}_k(\mathcal{D})\cap\mathcal{F}_k(\widetilde{\mathcal{D}})|$. False positive is defined to be the number of infrequent sequential patterns in $\mathcal{D}$ that are mistakenly included in $\mathcal{F}_k(\widetilde{\mathcal{D}})$, that is, $$|\mathcal{F}_k(\widetilde{\mathcal{D}})-\mathcal{F}_k(\mathcal{D})\cap\mathcal{F}_k(\widetilde{\mathcal{D}})|.$$ 
False drop is defined to be the number of frequent sequential patterns in $\mathcal{F}_k(\mathcal{D})$ that are wrongly omitted in $\mathcal{F}_k(\widetilde{\mathcal{D}})$, that is, $$|\mathcal{F}_k(\mathcal{D})\cup\mathcal{F}_k(\widetilde{\mathcal{D}})-\mathcal{F}_k(\widetilde{\mathcal{D}})|.$$
Since in our setting $|\mathcal{F}_k(\mathcal{D})| = |\mathcal{F}_k(\widetilde{\mathcal{D}})| = k$, false positives always equal false drops. 


\section{Sanitization Algorithm} \label{sec:sanitization}
We first provide an overview of our two-step sanitization algorithm in Algorithm~\ref{algo:sanitization}. Given a raw trajectory dataset $\mathcal{D}$, a privacy budget $\epsilon$, and a specified height of the prefix tree $h$, it returns a sanitized dataset $\widetilde{\mathcal{D}}$ satisfying $\epsilon$-differential privacy. \emph{BuildNoisyPrefixTree} builds a \emph{noisy} prefix tree $\mathcal{PT}$ for $\mathcal{D}$ using a set of count queries; \emph{GeneratePrivateRelease} employs a utility boosting technique on $\mathcal{PT}$ and then generates a differentially private release.

\begin{algorithm}[t]
{\bf Input:} Raw trajectory dataset $\mathcal{D}$ \\
{\bf Input:} Privacy budget $\epsilon$\\
{\bf Input:} Height of the prefix tree $h$\\
{\bf Output:} Sanitized dataset $\widetilde{\mathcal{D}}$

\begin{algorithmic}[1]
\small{
\STATE Noisy prefix tree $\mathcal{PT} \leftarrow BuildNoisyPrefixTree(\mathcal{D}, \epsilon, h)$;
\STATE Sanitized dataset $\widetilde{\mathcal{D}} \leftarrow GeneratePrivateRelease(\mathcal{PT})$;
\STATE {\bf return} $\widetilde{\mathcal{D}}$;}
\end{algorithmic}
\caption{Trajectory Data Sanitization Algorithm}\label{algo:sanitization}
\end{algorithm}

\subsection{Noisy Prefix Tree Construction}\label{subsec:noisyprefixtreeconstruction}

The noisy prefix tree of the raw trajectory dataset $\mathcal{D}$ cannot be simply generated at a time by scanning the dataset once, as the way we construct a deterministic prefix tree. To satisfy differential privacy, we need to guarantee that every trajectory that can be derived from the location universe (either in or not in $\mathcal{D}$) has certain probability to appear in the sanitized dataset so that the sensitive information in $\mathcal{D}$ could be masked. 

Our strategy for \emph{BuildNoisyPrefixTree} is to recursively group trajectories in $\mathcal{D}$ into \emph{disjoint} sub-datasets based on their prefixes and resort to the well-understood query model to guarantee differential privacy. Procedure~\ref{proc:buildprefixtree} presents the details of \emph{BuildNoisyPrefixTree}. We first create a prefix tree $\mathcal{PT}$ with a virtual root $Root(\mathcal{PT})$ (Lines 2-4). To build $\mathcal{PT}$, we employ a \emph{uniform} privacy budget allocation scheme, that is, divide the total privacy budget $\epsilon$ into equal portions $\bar{\epsilon} = \frac{\epsilon}{h}$, each is used for constructing a level of $\mathcal{PT}$ (Line 5). In Lines 6-19, we iteratively construct each level of $\mathcal{PT}$ in a noisy way. At each level, for each node $v$, we consider \emph{every} location in $\mathcal{L}$ as $v$'s potential child $u$ in order to satisfy differential privacy. Our goal is to identify the children that are associated with non-zero number of trajectories so that we can continue to expand them. However, we cannot make decision based on true numbers, but noisy counts. One important observation is that all such potential children are associated with \emph{disjoint} trajectory subsets and therefore $\bar{\epsilon}$ can be used \emph{in full} for each $u$ because of Theorem~\ref{thm:pcomposition}.

\begin{procedure}
{\bf Input:} Raw trajectory dataset $\mathcal{D}$ \\
{\bf Input:} Privacy budget $\epsilon$\\
{\bf Input:} Height of the prefix tree $h$\\
{\bf Output:} Noisy prefix tree $\mathcal{PT}$\\ [1ex]
\hspace*{2mm}1:  $i = 0$; \\
\hspace*{2mm}2:  Create an empty prefix tree $\mathcal{PT}$;\\
\hspace*{2mm}3:  Insert a virtual root $Root(\mathcal{PT})$ to $\mathcal{PT}$;\\
\hspace*{2mm}4:  Add all trajectories in $\mathcal{D}$ to $tr(Root(\mathcal{PT}))$; \\
\hspace*{2mm}5:  $\bar{\epsilon} = \frac{\epsilon}{h}$; \\
\hspace*{2mm}6:  {\bf while} $i < h$ {\bf do} \\
\hspace*{2mm}7:  \hspace{5mm}{\bf for} each node $v \in level(i, \mathcal{PT})$ {\bf do}\\
\hspace*{2mm}8:  \hspace{10mm}Generate a candidate set of nodes $\mathcal{U}$ from $\mathcal{L}$,\\
\hspace*{2mm} \hspace{12.8mm}each labeled by a location $L \in \mathcal{L}$;\\
\hspace*{2mm}9:  \hspace{10mm}{\bf for} each node $u \in \mathcal{U}$ {\bf do}\\ 
\hspace*{2mm}10:  \hspace{13.5mm}Consider $u$ as $v$'s child;\\
\hspace*{2mm}11:  \hspace{13.5mm}Add the trajectories $D$ in $tr(v)$ s.t. \\
\hspace*{2mm} \hspace{17.7mm}$prefix(u, \mathcal{PT}) \preceq D$ to $tr(u)$;\\
\hspace*{2mm}12:  \hspace{13.5mm}$c(u) = NoisyCount(|tr(u)|, \bar{\epsilon})$; \\
\hspace*{2mm}13:  \hspace{13.5mm}{\bf if} $c(u) \geq \theta$ {\bf then}\\
\hspace*{2mm}14:  \hspace{18.5mm}Add $u$ to $\mathcal{PT}$ as $v$'s child; \\
\hspace*{2mm}15:  \hspace{13.5mm}{\bf end if}\\
\hspace*{2mm}16:  \hspace{8.5mm}{\bf end for}\\
\hspace*{2mm}17:  \hspace{3.5mm}{\bf end for}\\
\hspace*{2mm}18:  \hspace{3.5mm}$i++$;\\
\hspace*{2mm}19:  {\bf end while}\\
\hspace*{2mm}20:  {\bf return} $\mathcal{PT}$; \\
  \caption{$BuildNoisyPrefixTree$ Procedure} \label{proc:buildprefixtree}
\end{procedure}

For a dataset with a very large location universe $\mathcal{L}$, processing all locations explicitly may be slow. We provide an efficient implementation by \emph{separately} handling potential child nodes associated with non-zero and zero number of trajectories (referred as \emph{non-empty node} and \emph{empty node} respectively in the following). For a non-empty node $u$, we add Laplace noise to $|tr(u)|$ and use the noisy answer $c(u)=NoisyCount(|tr(u)|, \bar{\epsilon})$ to decide if it is non-empty. If $c(u)$ is greater than or equal to the threshold $\theta = \frac{2\sqrt{2}}{\bar{\epsilon}}$ (two times of the standard deviation of noise), we deem that $u$ is ``non-empty" and insert $u$ to $\mathcal{PT}$ as $v$'s child. We choose a relatively large threshold mainly for the reason of efficiency, and meanwhile it also has a positive impact on utility because more noisy nodes can be pruned. Since non-empty nodes are typically of a small number, this process can be done efficiently.

For the empty nodes, we need to conduct a series of \emph{independent} boolean tests, each calculates $NoisyCount(0, \bar{\epsilon})$ to check if it passes $\theta$. The number of empty nodes that pass $\theta$, $k$, follows the \emph{binomial distribution} $B(m, p_{\theta})$, where $m$ is the total number of empty nodes we need to check and $p_{\theta}$ is the probability for a single experiment to succeed. Inspired by Cormode et al.'s work~\cite{CPS11}, we design a statistical process for Laplace mechanism to directly extract $k$ empty nodes without explicitly processing every empty node  (in~\cite{CPS11}, Cormode et al. design a statistical process for \emph{geometric mechanism}~\cite{GRS09}, a discretized version of Laplace mechanism).

\begin{figure*}[t]
\centering
\scalebox{0.8}{
\includegraphics[bb= 41 29 566 145]{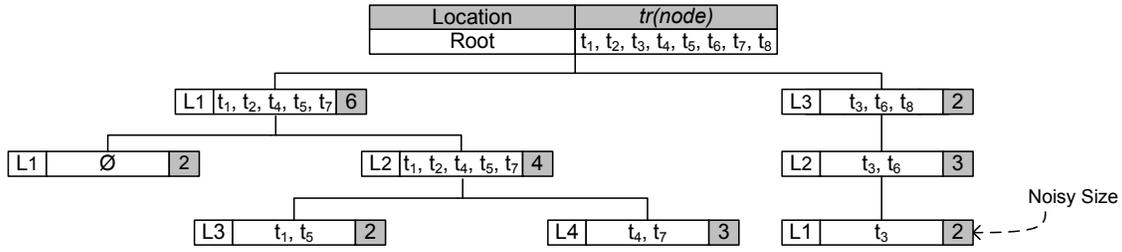}
}
\caption{The noisy prefix tree of the sample data}\label{figure:noisyprefixtree}
\end{figure*}

\vspace{2mm}
\bt \label{thm:zerocountextraction} \em Independently conducting $m$ pass/not pass experiments based on Laplace mechanism with privacy budget $\bar{\epsilon}$ and threshold $\theta$ is equivalent to the following steps: 
\begin{enumerate}
	\item Generate a value $k$ from the binomial distribution $B(m, p_{\theta})$, where $p_{\theta}= \frac{exp(-\bar{\epsilon} \theta)}{2}$.
	\item Select $k$ uniformly random empty nodes without replacement with noisy counts sampled from the distribution 
	$$P(x) = \begin{cases} 0 & \forall x < \theta \\ 1- exp(\bar{\epsilon}\theta-\bar{\epsilon}x)& \forall x \geq \theta \end{cases}~\blob$$
\end{enumerate}
\et 
\vspace{2mm}
\emph{Proof.} The probability of a single experiment passing the threshold $\theta$ is $$Pr[PASS] = \int_{\theta}^{\infty}\frac{\bar{\epsilon}}{2}exp(-\bar{\epsilon} x)dx = \frac{exp(-\bar{\epsilon} \theta)}{2}.$$ Since the experiments are independent, the number of successful experiments, $k$, follows the binomial distribution $B(m, \frac{exp(-\bar{\epsilon} \theta)}{2})$. Once $k$ is determined, we can uniformly at random select $k$ empty nodes. The probability density function of the noisy counts $x$ for the $k$ empty nodes, conditional on $x \geq \theta$, is:
$$p(x|x \geq \theta) = \begin{cases} 0 & \forall x < \theta \\ \frac{\frac{\bar{\epsilon}}{2}exp(-\bar{\epsilon}x)}{p_{\theta}} = \bar{\epsilon}exp(\bar{\epsilon}\theta - \bar{\epsilon}x) & \forall x \geq \theta \end{cases}$$

The corresponding cumulative distribution function is:
$$P(x) = \begin{cases} 0 & \forall x < \theta \\ \int_{\theta}^{x}\bar{\epsilon}exp(\bar{\epsilon}\theta - \bar{\epsilon}x)dx = 1- exp(\bar{\epsilon}\theta-\bar{\epsilon}x)& \forall x \geq \theta \end{cases}$$

This completes the proof.~\blob
\vspace{2mm}

\noindent \textbf{Selection of $h$ value.} We discuss the selection of $h$ value. A natural choice of $h$ value is the maximum trajectory length of $\mathcal{D}$. However, this choice raises two problems. First, for most trajectory databases, the maximum trajectory length is much longer than the average trajectory length (see Section~\ref{subsec:dataset} for two examples), implying that extremely long trajectories are of very small supports. Keeping expanding the prefix tree for such long trajectories is counter-productive: it ends up with many noisy trajectories that do not exist in the original database and therefore results in poor utility. Second, in some cases, the maximum trajectory length itself may be sensitive. Since the length of a trajectory in $\mathcal{D}$ could be arbitrarily large, the sensitivity of maximum trajectory length is \emph{unbounded}. It follows that it is very difficult to design a differentially private mechanism to obtain a reliable maximum trajectory length.  

Evfimievski et al.~\cite{ESAG04} provides an insightful observation that could be used for the selection of $h$ value. They point out that, in the context of transaction data, it is typically impossible to make transactions of size 10 and longer both privacy preserving and useful (regardless of the underlying dataset). This observation also applies to trajectory data. Long trajectories carry too much sensitive information to achieve a reasonable trade-off between privacy and utility. Theoretically, selecting a relatively small $h$ value has a limited negative effect on the resulting utility. For count queries, long trajectories have very small supports. Moreover, Procedure~\ref{proc:buildprefixtree} does not eliminate entire long trajectories, but just truncates their tailing parts, and therefore results in even smaller relative errors. Similarly, for frequent sequential pattern mining, the elimination of tailing parts of long trajectories with small supports does not significantly affect resulting frequent sequential patterns, which are usually of a small length. We experimentally confirm our analysis in Section~\ref{sec:experiment}.

\vspace{2mm}
\bx \label{ex:prefixtreeconstruction} \em Given the trajectory database $\mathcal{D}$ in Table~\ref{table:rawdata}, the height $h=3$, and the calculated threshold $\theta = 2$, the construction of a possible noisy prefix tree $\mathcal{PT}$ is illustrated in Figure~\ref{figure:noisyprefixtree}. A path of $\mathcal{PT}$ may be of a length shorter than $h$ if it has been considered ``empty" before $h$ is reached.~\blob
\ex

\subsection{Private Release Generation}
Based on the noisy prefix tree $\mathcal{PT}$, we can generate the sanitized database by traversing $\mathcal{PT}$ once in \emph{postorder}, calculating the number $n$ of trajectories terminated at each node $v$ and appending $n$ copies of $prefix(v, \mathcal{PT})$ to the output. However, due to the noise added to ensure differential privacy, we may not be able to obtain a meaningful release. For example, in Figure~\ref{figure:noisyprefixtree}, consider the root-to-leaf path $Root(\mathcal{PT}) \ra L_3 \ra L_2 \ra L_1$. We have $c(L_2) > c(L_3)$, which is counterintuitive because it is not possible, in general, to have more trajectories with the prefix $prefix(u, \mathcal{PT})$ than trajectories with the prefix $prefix(v, \mathcal{PT})$, where $u$ is a child of $v$ in $\mathcal{PT}$. If we leave such inconsistencies unsolved, the resulting release may not be meaningful and therefore provides poor utility. 

\vspace{2mm}
\bde[Consistency Constraint]  \label{dfn:constraint} \em In a non-noisy prefix tree, there exist two sets of consistency constraints: 
\begin{enumerate}
	\item For any root-to-leaf path $p$, $\forall v_i \in p, |tr(v_{i})| \leq |tr(v_{i+1})|$, where $v_i$ is a child of $v_{i+1}$;
	\item For each node $v$, $|tr(v)| \geq \sum_{u \in children(v)}{|tr(u)|}$.~\blob
\end{enumerate} \ede 
\vspace{2mm}

Our goal is to enforce such consistency constraints on the noisy prefix tree in order to produce a consistent and more accurate private release. We adapt the constrained inference technique proposed in~\cite{HRMS10} to adjust the noisy counts of nodes in the noisy prefix tree so that the constraints defined in Definition~\ref{dfn:constraint} are respected. Note that the technique proposed in~\cite{HRMS10} cannot be directly applied to our case because: 1) the noisy prefix tree has an irregular structure (rather than a complete tree with a fixed degree); 2) the noisy prefix tree has different constraints $|tr(v)| \geq \sum_{u \in children(v)}{|tr(u)|}$ (rather than $|tr(v)| = \sum_{u \in children(v)}{|tr(u)|}$). Consequently, we propose a two-phase procedure to obtain a \emph{consistent} estimate with respect to Definition~\ref{dfn:constraint} for each node (except the virtual root) in the noisy prefix tree $\mathcal{PT}$. 

We first generate an intermediate estimate for the noisy count of each node $v$ (except the virtual root) in $\mathcal{PT}$. Consider a root-to-leaf path $p$ of $\mathcal{PT}$. Let us organize the noisy counts of nodes $v_i \in p$ into a sequence $\mathcal{S} = \langle c(v_1), c(v_2), \cdots, c(v_{|p|}) \rangle$, where $v_{i}$ is a child of $v_{i+1}$. Let $mean[i, j]$ denote the mean of a subsequence of $\mathcal{S}$, $\langle c(v_i), c(v_{i+1}), ..., c(v_j) \rangle$, that is, $mean[i, j] = \frac{\sum_{m=i}^{j}{c(v_m)}}{j-i+1}$. We compute the intermediate estimates $\widetilde{\mathcal{S}}$ by Theorem~\ref{thm:boost}~\cite{HRMS10}.

\vspace{3mm}
\bt \label{thm:boost} \em Let $L_m = min_{j \in [m, n]}max_{i \in [1, j]}mean[i, j]$ and $U_m = max_{i \in [1, m]}min_{j \in [i, |p|]}mean[i,j]$. $\widetilde{\mathcal{S}} = \langle L_1, L_2, ..., L_{|p|} \rangle = \langle U_1, U_2, ..., U_{|p|} \rangle $.~\blob \et 
\vspace{3mm}

The result of Theorem~\ref{thm:boost} is the \emph{minimum $L_2$ solution}~\cite{HRMS10} of $\mathcal{S}$ that satisfies the first type of constraints in Definition~\ref{dfn:constraint}. However, a node $v$ in $\mathcal{PT}$ appears in $|leaves(v, \mathcal{PT})|$ root-to-leaf paths, where $leaves(v, \mathcal{PT})$ denotes the leaves of the subtree of $\mathcal{PT}$ rooted at $v$, and therefore, has $|leaves(v, \mathcal{PT})|$ intermediate estimates, each being an independent observation of the true count $|tr(v)|$. We compute the \emph{consolidated} intermediate estimate of $v$ as the mean of the estimates, normally the best estimate for $|tr(v)|$~\cite{JRT97}. We denote the consolidated intermediate estimate of $v$ by $\widetilde{c}(v)$. 

After obtaining $\widetilde{c}(v)$ for each node $v$, we compute its consistent estimate $\bar{c}(v)$ in a top-down fashion as follows: 
\vspace{1mm}
$$\bar{c}(v) = \begin{cases} \widetilde{c}(v) \ \ \ \ \ \ \ \ \ \ \ \ \ \ \ \ \ \ \ \ \ \ \ \ \ \ \ \ \ \ \ \ \ if v\in {level(1, \mathcal{PT})} \\ \widetilde{c}(v) + min(0, \frac{\bar{c}(w)-\sum_{u\in children(w)}\widetilde{c}(u)}{|children(w)|}) \ \ otherwise \end{cases}$$
\vspace{1mm}

\noindent where $w$ is the parent of $v$. It follows the intuition that the observation $\sum_{u\in children(w)}\widetilde{c}(u) > \bar{c}(w)$ is strong evidence that excessive noise is added to the children. Since the magnitude of noise in $\bar{c}(w)$ is approximately $|children(w)|$ times smaller than $\sum_{u\in children(w)}\widetilde{c}(u)$, it is reasonable to decrease the children's counts according to $\bar{c}(w)$. However, we never increase the children's counts based on $\bar{c}(w)$ because a large $\bar{c}(w)$ simply indicates that many trajectories actually terminate at $w$. It is easy to see that the consistency constraints in Definition~\ref{dfn:constraint} are respected among consistent estimates, and therefore the proof is omitted here.

Once we obtain the consistent estimate for each node, we can generate the private release by a postorder traversal of $\mathcal{PT}$. In Section~\ref{sec:experiment}, we demonstrate that the constrained inferences significantly improve utility of sanitized data.

\subsection{Analysis}\label{subsec:analysis}
\noindent\textbf{Privacy Analysis.} Kifer and Machanavajjhala~\cite{KM11} point out that differential privacy must be applied with caution. The privacy protection provided by differential privacy relates to the data generating mechanism and deterministic aggregate-level background knowledge. In most cases, for example the STM case, the trajectories in the raw database are independent of each other and therefore \emph{the evidence of participation}~\cite{KM11} of a record owner can be eliminated by removing his record. Furthermore, we assume that no deterministic statistics of the raw database will ever be released. Hence differential privacy is appropriate for our problem. We now show that Algorithm~\ref{algo:sanitization} satisfies $\epsilon$-differential privacy.

\vspace{2mm}
\bt \label{thm:dp} \em Given the total privacy budget $\epsilon$, Algorithm~\ref{algo:sanitization} ensures $\epsilon$-differential privacy.~\blob \et 
\vspace{2mm}

\emph{Proof.} Algorithm~\ref{algo:sanitization} consists of two steps, namely \emph{BuildNoisyPrefixTree} and \emph{GeneratePrivateRelease}. In the procedure \emph{BuildNoisyPrefixTree}, our approach appeals to the well-understood query model to construct the noisy prefix tree $\mathcal{PT}$. Consider a level of $\mathcal{PT}$. Since all nodes on the same level contain \emph{disjoint} sets of trajectories. According to Theorem~\ref{thm:pcomposition}, the entire privacy budget needed for a level is bounded by the worst case, that is, $\bar{\epsilon} = \frac{\epsilon}{h}$. The use of privacy budget on different levels follows Theorem~\ref{thm:scomposition}. Since there are at most $h$ levels, the total privacy budget needed to build the noisy prefix tree is $\leq h \times \bar{\epsilon} = \epsilon$.

For the procedure \emph{GeneratePrivateRelease}, we make use of the inherent constraints of a prefix tree to boost utility. The procedure only accesses a differentially private noisy prefix tree, not the underlying database. As proven by Hay et al.~\cite{HRMS10}, a post-processing of differentially private results remains differentially private. Therefore, Algorithm~\ref{algo:sanitization} as a whole maintains $\epsilon$-differential privacy.~\blob

\vspace{2mm}
\noindent\textbf{Complexity Analysis.} Algorithm~\ref{algo:sanitization} is of runtime complexity $O(|\mathcal{D}|\cdot|\mathcal{L}|)$, where $|\mathcal{D}|$ is the number of trajectories in the input database $\mathcal{D}$ and $|\mathcal{L}|$ is the size of the location universe. It comes from the following facts. For \emph{BuildNoisyPrefixTree}, the major computational cost is node generation and trajectory distribution. For each level of the noisy prefix tree, the number of nodes to generate approximates $k|\mathcal{D}|$, where $k \ll |\mathcal{L}|$ is a number depending on $|\mathcal{L}|$. For each level, we need to distribute \emph{at most} $|\mathcal{D}|$ trajectories to the newly generated nodes. Hence, the complexity of constructing a single level is $O(|\mathcal{D}|\cdot|\mathcal{L}|)$. Therefore, the total runtime complexity of \emph{BuildNoisyPrefixTree} for constructing a noisy prefix tree of height $h$ is $O(h|\mathcal{D}|\cdot|\mathcal{L}|)$. In \emph{GeneratePrivateRelease}, the complexity of calculating the intermediate estimates for a single root-to-leaf path is $O(h^2)$. Since there can be at most $|\mathcal{D}|$ distinct root-to-leaf paths, the complexity of computing all intermediate estimates is $O(h^2|\mathcal{D}|)$. To compute consistent estimates, we need to visit every node exactly twice, which is of complexity $O(|\mathcal{D}|\cdot|\mathcal{L}|)$. Similarly, the computational cost of generating the private release by traversing the noisy prefix tree once in postorder is $O(|\mathcal{D}|\cdot|\mathcal{L}|)$. Since $h$ is a very small constant compared to $|\mathcal{D}|$ and $|\mathcal{L}|$, the total complexity of Algorithm~\ref{algo:sanitization} is $O(|\mathcal{D}|\cdot|\mathcal{L}|)$. 

\subsection{Extensions} \label{subsec:extensibility} 
In some applications, trajectory data may have locations associated with \emph{timestamps}. The time factor is often discretized into intervals at different levels of granularity, e.g., hour, which is typically determined by the data publisher. All timestamps of a trajectory database form a timestamp universe. This type of trajectories is composed of a sequence of \emph{location}-\emph{timestamp} pairs in the form of $loc_1t_1 \ra loc_2t_2 \ra \cdots \ra loc_nt_n$, where $t_1 \leq t_2 \leq \cdots \leq t_n$. 

Our solution can be seamlessly applied to this type of trajectory data. In this case, we can label each node in the prefix tree by both a location and a timestamp. Therefore, two trajectories with the same sequence of locations but different timestamps are considered different. For example, $L_1T_1 \ra L_2T_2$ is different from $L_1T_2 \ra L_2T_3$, and the corresponding prefix tree will have two non-overlapping root-to-leaf paths. Consequently, when constructing the noisy prefix tree, in order to expand a node $loc_it_i$, we have to consider the combinations of all locations and the timestamps in the time universe that are greater than $t_i$ (because the timestamps in a trajectory are non-decreasing), resulting in a larger candidate set. Due to the efficient implementation we propose in Section~\ref{subsec:noisyprefixtreeconstruction}, the computational cost will remain moderate.   


\section{Experimental Evaluation} \label{sec:experiment}
In this section, our objective is to examine the utility of sanitized data in terms of count queries and frequent sequential pattern mining, and to evaluate the scalability of our approach for processing large-scale real-life trajectory data. In particular, we study the utility improvements due to constrained inferences. In the following, we refer the approach without constrained inferences as \emph{Basic} and the one with constrained inferences as \emph{Full}. We are not able to compare our solution to others because there does not exist any other solution that is able to process large-volume trajectory data in the framework of differential privacy. Our implementation was done in C++, and all experiments were performed on an Intel Core 2 Duo 2.26GHz PC with 2GB RAM.

\subsection{Dataset} \label{subsec:dataset}

\begin{figure}[t]
\centering
\scalebox{0.43}{
\includegraphics[bb= 54 292 536 580]{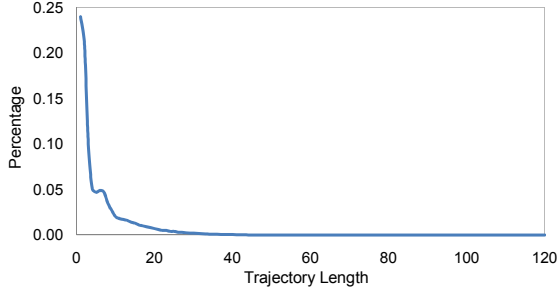}
}
\caption{The distribution of trajectory length in the \emph{STM} dataset}\label{figure:distribution}
\end{figure}

We perform extensive experiments on the real-life trajectory dataset \emph{STM}, which is provided by the \emph{Soci\'{e}t\'{e} de transport de Montr\'{e}al} (STM). The \emph{STM} dataset records the transit history of passengers in the STM transportation network. Each trajectory in the dataset is composed of a time-ordered list of stations visited by a passenger within a week in 2009. The \emph{STM} dataset contains $1,210,096$ records and has a location universe of size $1,012$. The maximum trajectory length $max|D|$ of the \emph{STM} dataset is $121$ and the average length $avg|D|$ is 6.7. We show the distribution of trajectory length of the \emph{STM} dataset in Figure~\ref{figure:distribution}, which justifies our rationale of selecting $h$ value in Section~\ref{subsec:noisyprefixtreeconstruction}. Note that similar observations can be found in many real-life trajectory datasets, for example, the \emph{MSNBC} dataset~\footnote{\emph{MSNBC} is publicly available at UCI machine learning repository (http://archive.ics.uci.edu/ml/index.html).} has the maximum trajectory length that is over 2500 times larger than the average trajectory length.



\subsection{Utility}

\noindent\textbf{Count Query}. In our first set of experiments, we examine relative errors of count queries with respect to two different parameters, namely the privacy budget $\epsilon$ and the noisy prefix tree height $h$. We follow the evaluation scheme from previous works~\cite{XWG10}, \cite{CMFDX11}. For each privacy budget, we generate 40,000 random count queries with varying numbers of locations. We call the number of locations in a query the \emph{length} of the query. We divide the query set into 4 subsets such that the query length of the $i$-th subset is uniformly distributed in $[1, \frac{ih}{4}]$ and each location is randomly drawn from the location universe $\mathcal{L}$. The sanity bound $s$ is set to $0.1\%$ of the dataset size, the same as~\cite{XWG10}, \cite{CMFDX11}.

\begin{figure}[t]
\centering $\begin{array}{c c}
    \multicolumn{1}{l}{\mbox{\bf }} & \multicolumn{1}{l}{\mbox{\bf }} \\
    \hspace{-3mm} \scalebox{0.27}{\includegraphics[bb= 54 202 524 630]{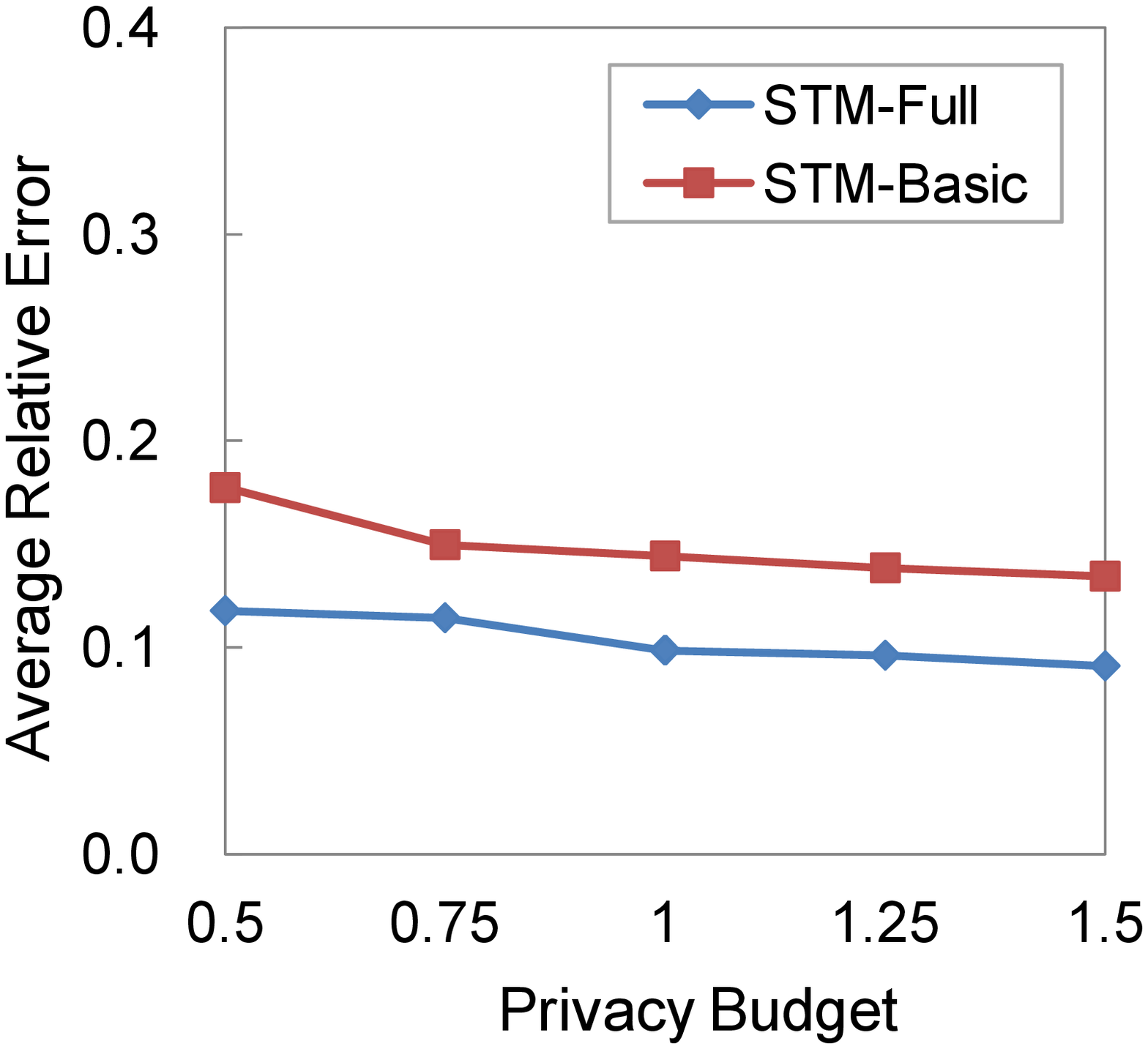}} & \hspace{-3mm} \scalebox{0.27}{\includegraphics[bb= 54 202 524 630]{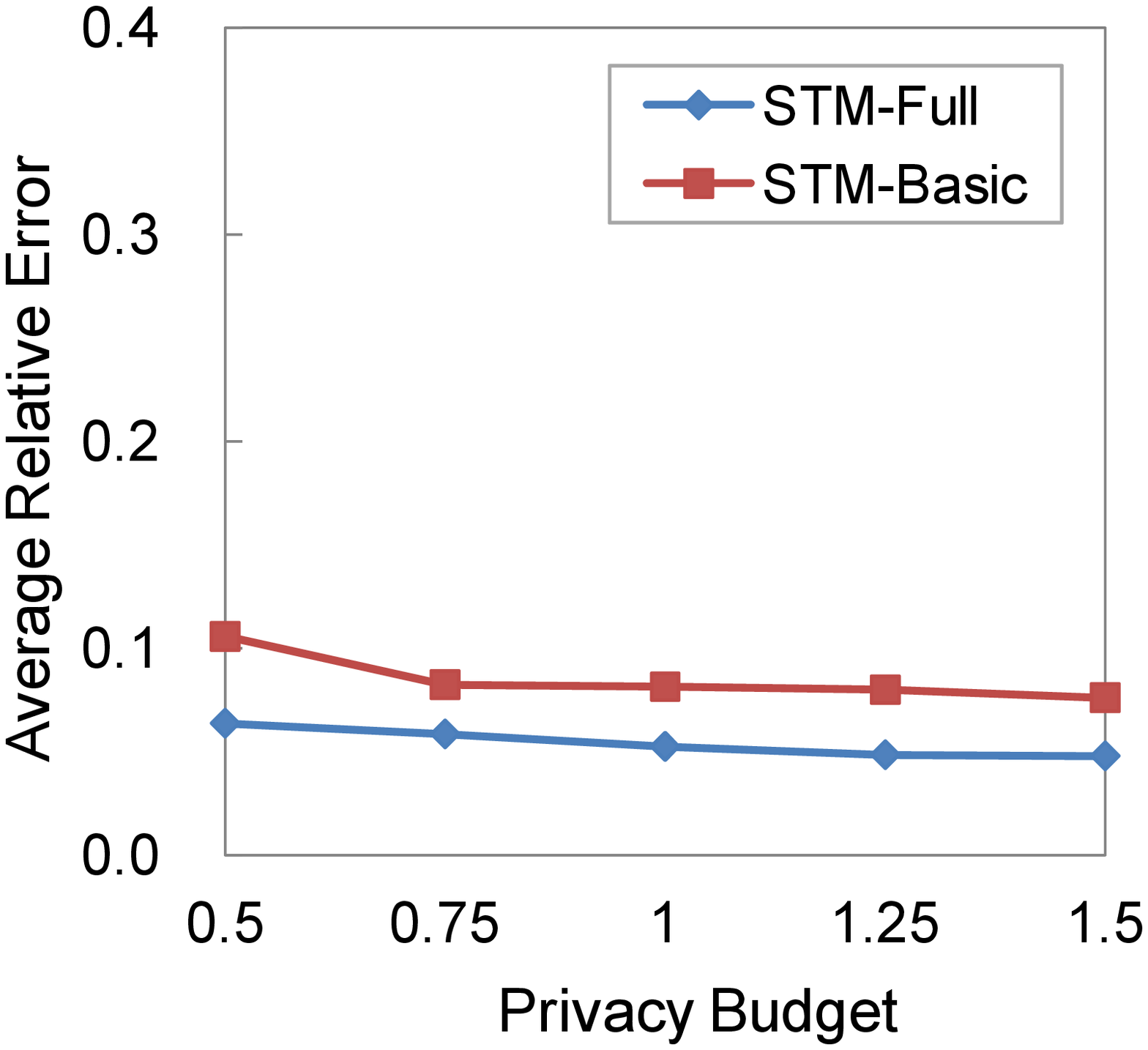}} \\
    \mbox{\small{(a) $max|Q|=3$}} & \mbox{\small{(b) $max|Q|=6$}} \\ 
    \multicolumn{1}{l}{\mbox{\bf }} & \multicolumn{1}{l}{\mbox{\bf }} \\
    \hspace{-3mm} \scalebox{0.27}{\includegraphics[bb= 54 202 524 630]{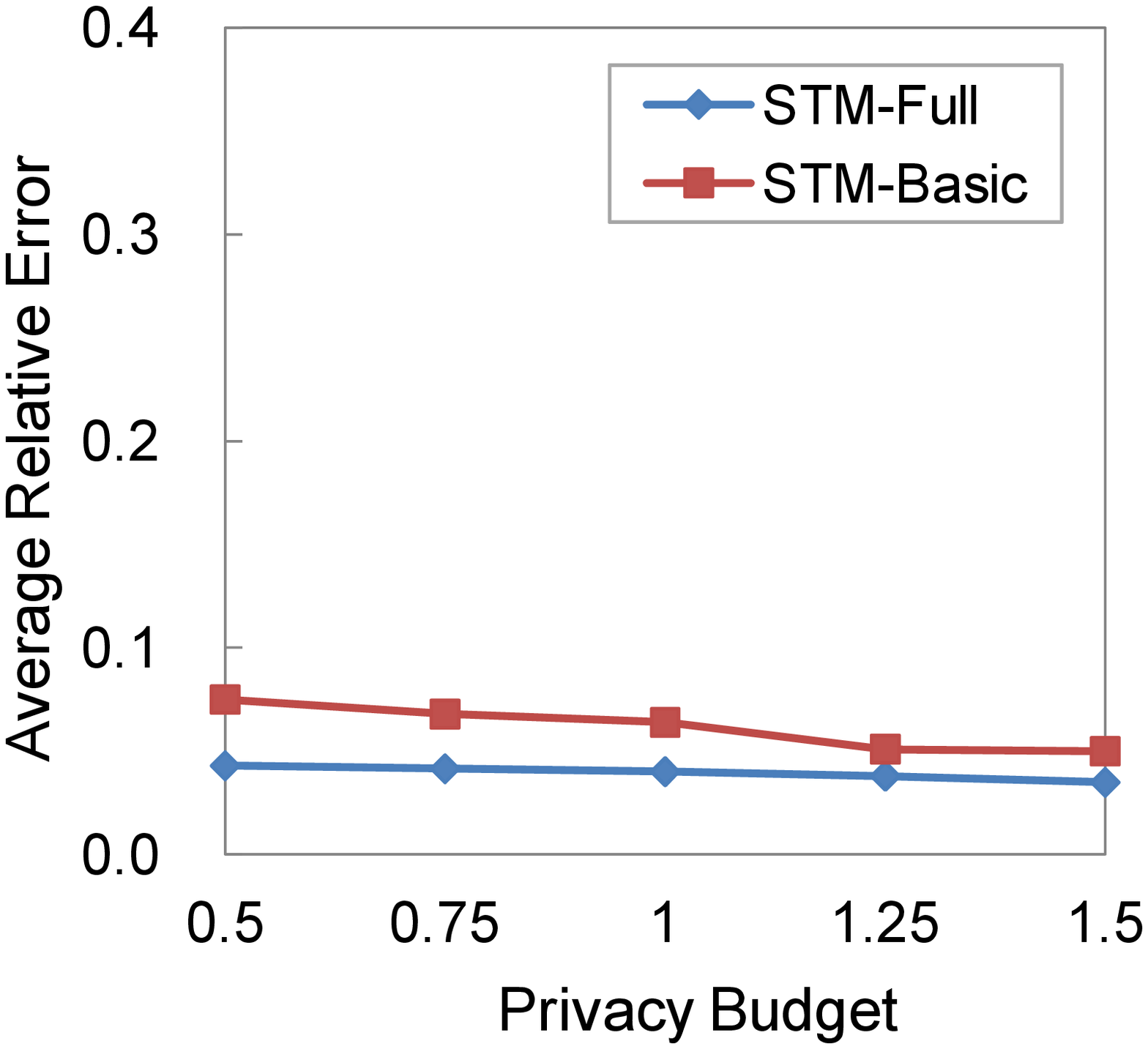}} & \hspace{-3mm} \scalebox{0.27}{\includegraphics[bb= 54 202 524 630]{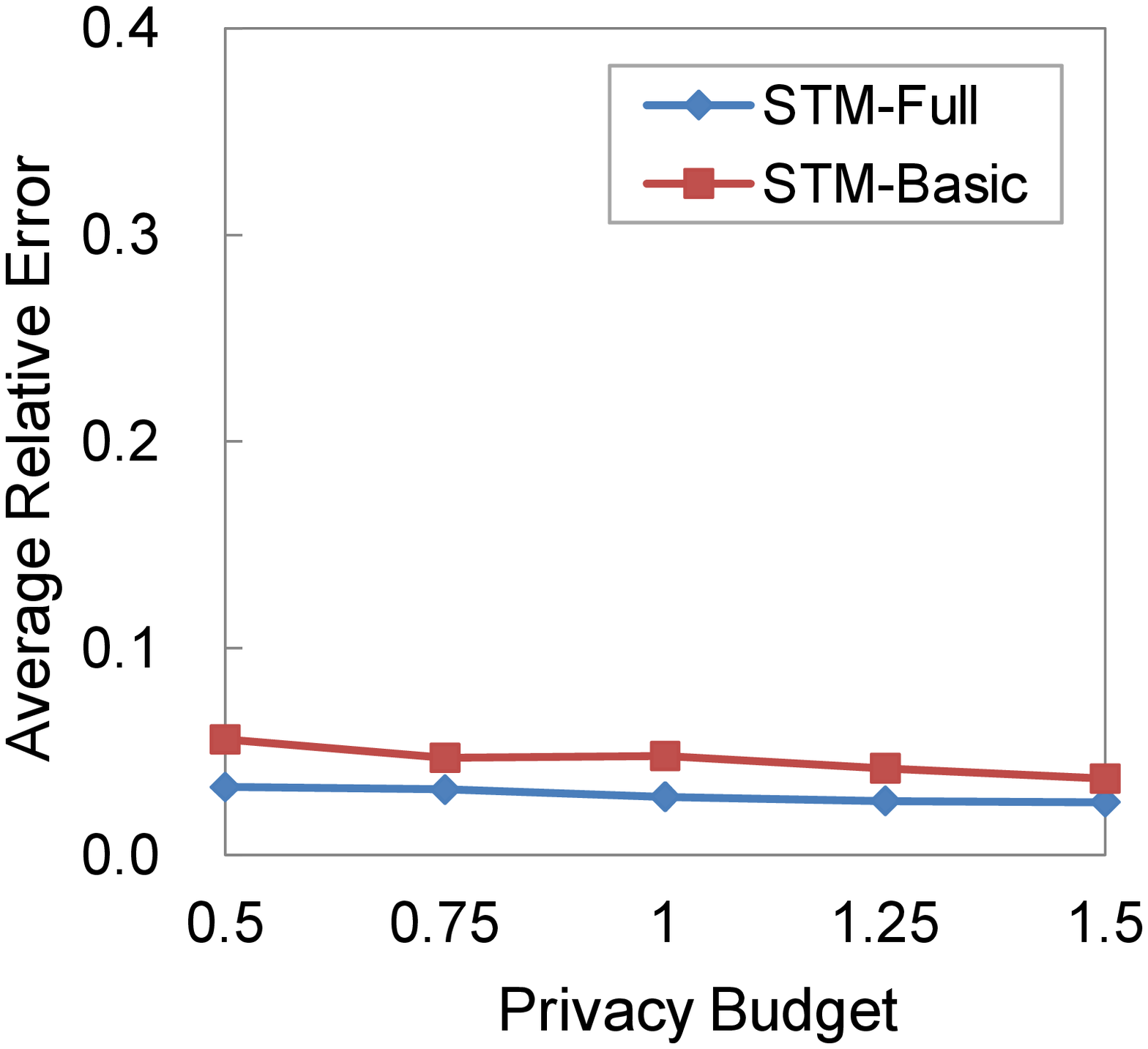}}\\ [0.05cm]
    \mbox{\small{(c) $max|Q|=9$}} & \mbox{\small{(d) $max|Q|=12$}} \\
    \end{array}$
\caption{Average relative error vs. privacy budget.}\label{figure:privacybudget}
\end{figure}

Figure~\ref{figure:privacybudget} examines the average relative errors under varying privacy budgets from 0.5 to 1.5 while fixing the noisy prefix tree height $h$ to 12. The X-axes represent the different subsets by their maximum query length $max|Q|$. As expected, the average relative error decreases when the privacy budget increases because less noise is added and the construction process is more precise. In general, our approach maintains high utility for count queries. Even in the worst case ($\epsilon = 0.5$ and $max|Q|=3$), the average relative error of \emph{Full} is still less than $12\%$. Under the typical setting of the privacy budget $\epsilon = 1.0$, the relative error is less than $10\%$ for all query subsets. Such level of relative error is deemed acceptable for data analysis tasks at the STM. The small relative error also demonstrates the distinct advantage of a data-dependent approach: noise would never accumulate exponentially as a data-independent approach.

Figure~\ref{figure:privacybudget} also shows that the constrained inference technique significantly reduces the average relative errors with $30\%-74\%$ improvement. The constrained inference technique is beneficial especially when the privacy budget is small.

\begin{figure}[ht]
\centering $\begin{array}{c c}
    \multicolumn{1}{l}{\mbox{\bf }} & \multicolumn{1}{l}{\mbox{\bf }} \\
    \hspace{-3mm} \scalebox{0.27}{\includegraphics[bb= 54 202 524 630]{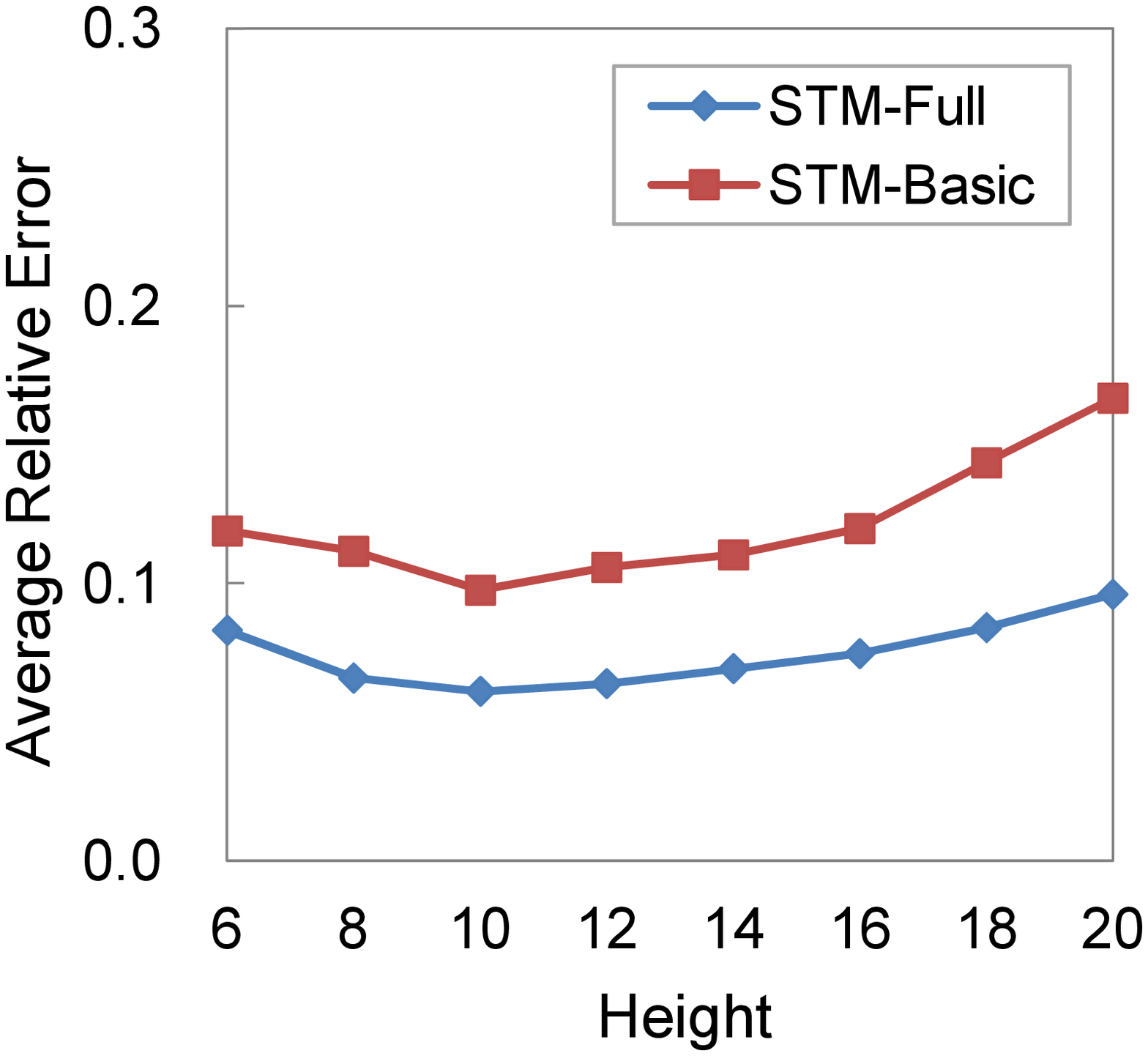}} & \hspace{-3mm} \scalebox{0.27}{\includegraphics[bb= 54 202 524 630]{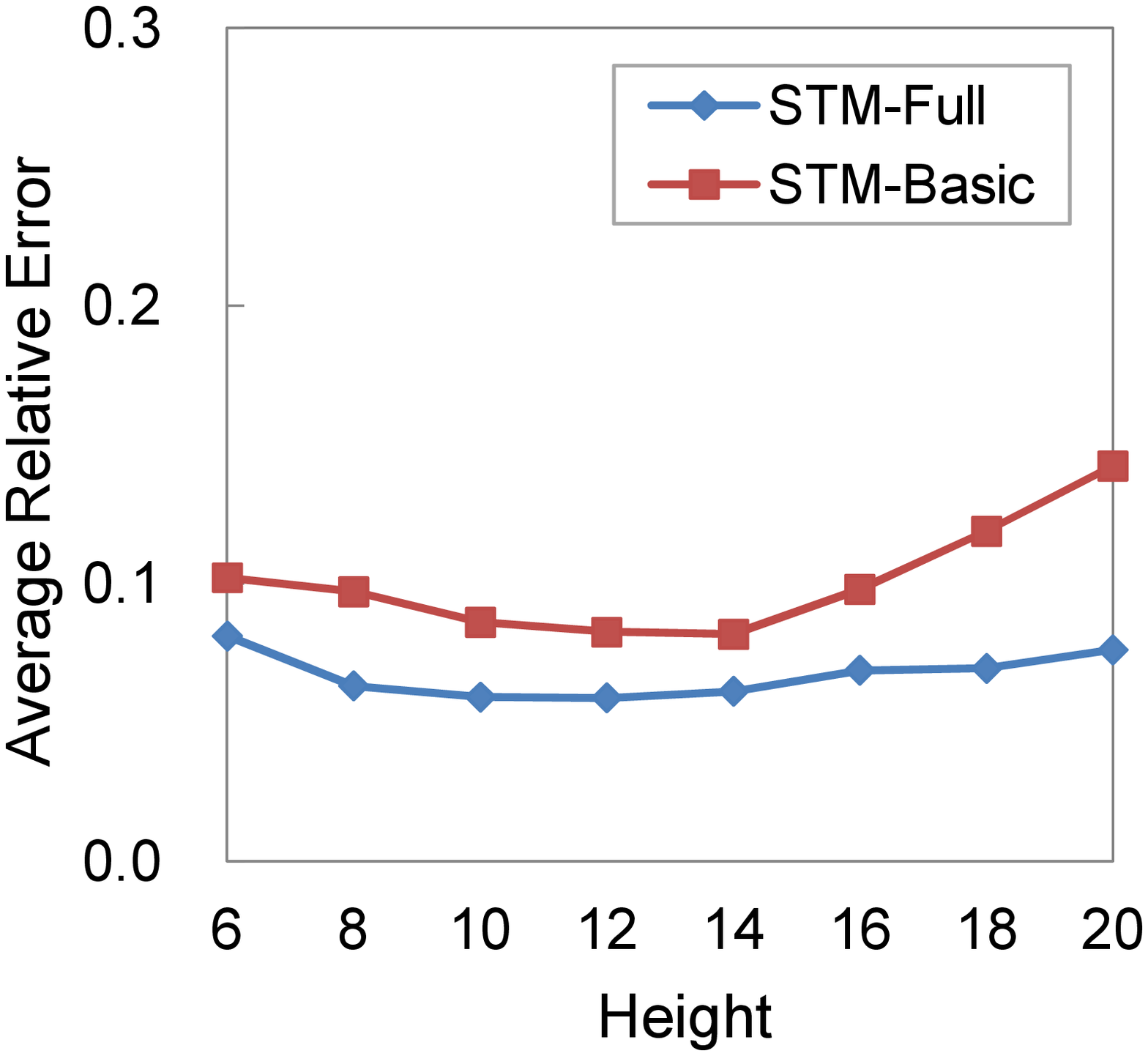}} \\
    \mbox{\small{(a) $B=0.5$}} & \mbox{\small{(b) $B=0.75$}} \\
    \multicolumn{1}{l}{\mbox{\bf }} & \multicolumn{1}{l}{\mbox{\bf }} \\
    \hspace{-3mm} \scalebox{0.27}{\includegraphics[bb= 54 202 524 630]{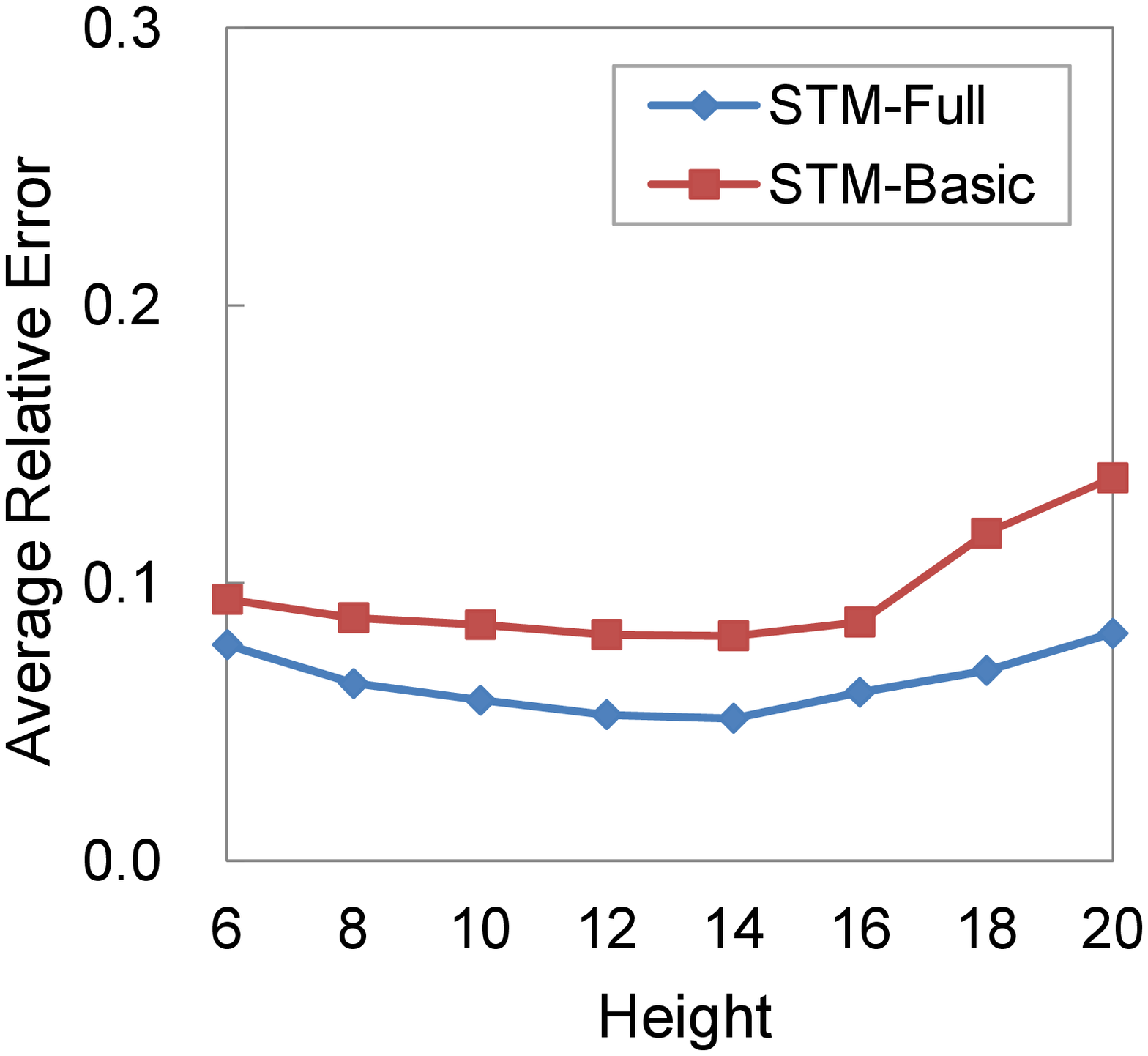}} & \hspace{-3mm} \scalebox{0.27}{\includegraphics[bb= 54 202 524 630]{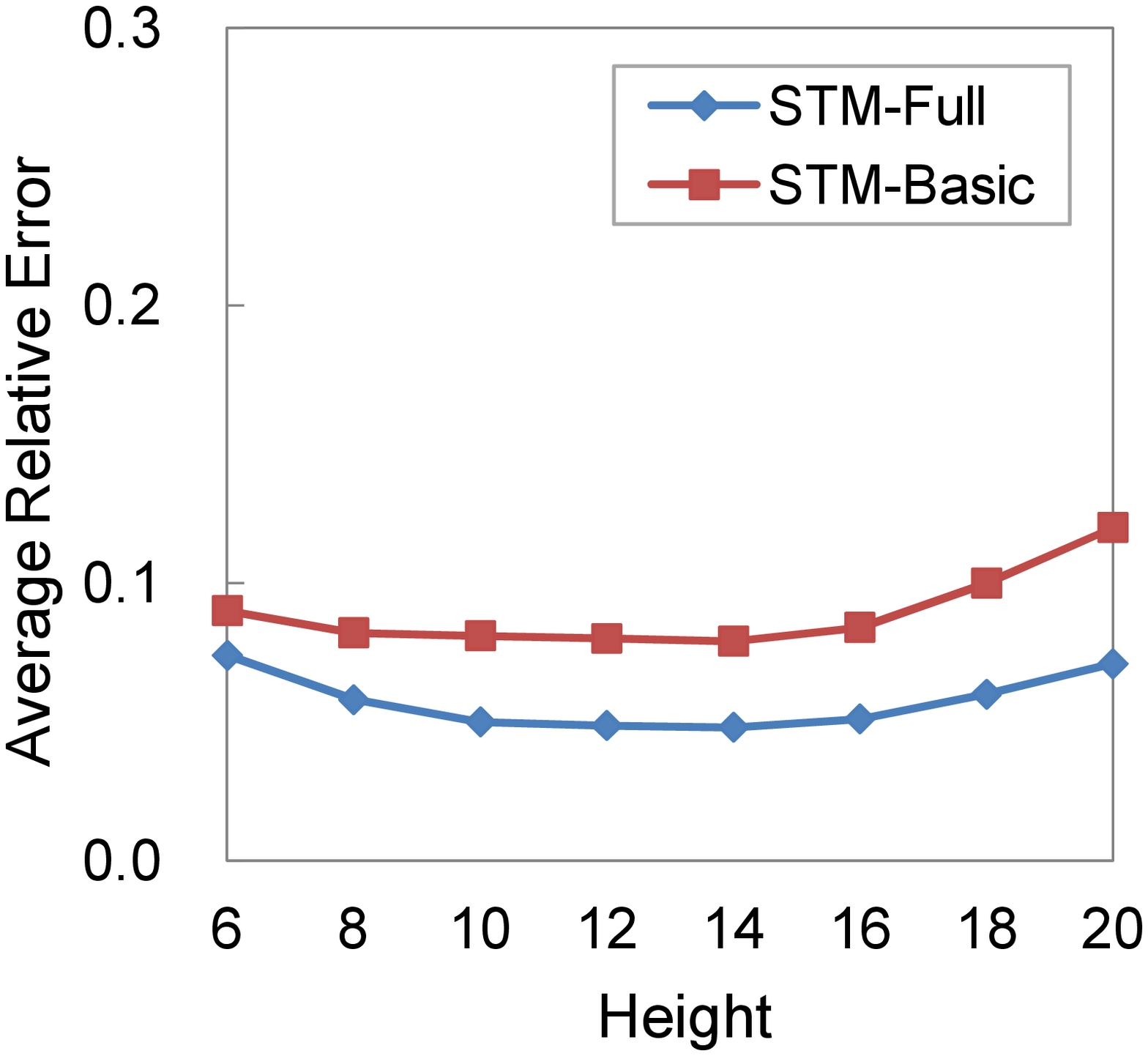}}\\ [0.05cm]
    \mbox{\small{(c) $B=1.0$}} & \mbox{\small{(d) $B=1.25$}} \\
    \end{array}$
\caption{Average relative error vs. prefix tree height.}\label{figure:height}
\end{figure}

Figure~\ref{figure:height} studies how the average relative errors vary under different noisy prefix tree height $h$ with query length fixed to 6. We can observe that with the increase of $h$, the relative errors do not decrease monotonically. Initially, the relative error decreases when $h$ increases because the increment of $h$ allows to retain more information from the underlying database. However, after a certain threshold, the relative error becomes larger with the increase of $h$. The reason is that when $h$ gets larger, the privacy budget assigned to each level becomes smaller, and therefore noise becomes larger. It is interesting to point out that in fact a larger $h$ value does not always lead to a noisy prefix tree with a larger height. Recall that a path of a noisy prefix tree needs to have sufficient number of trajectories to reach $h$ levels. As a result, specifying a larger $h$ may not be able to obtain more information from the underlying database, but only decreases the privacy budget of each level. This confirms our analysis in Section~\ref{subsec:noisyprefixtreeconstruction} that in practice a good $h$ value does not need to be very large. Moreover, from Figure~\ref{figure:height}, we can learn that desirable relative errors can be achieved within a relatively wide range of $h$ values (e.g., 10-16). Hence, it becomes easy for a data publisher to select a good $h$ value.

In addition, from Figure~\ref{figure:height}, we can see that the benefits of the constrained inference technique are two-fold. First, we obtain much smaller relative errors in all different settings. Second, the relative errors become more stable with respect to varying height values. 

\noindent\textbf{Frequent Sequential Pattern Mining}.
In the second set of experiments, we demonstrate the utility of sanitized data by a concrete data mining task, frequent sequential pattern mining. Specifically, we employ PrefixSpan to mine frequent sequential patterns~\footnote{An implementation of the PrefixSpan algorithm proposed in~\cite{PHMPCDH01}, available at http://code.google.com/p/prefixspan/.}. 

Table~\ref{table:support} shows how the utility changes with different top $k$ values while fixing $\epsilon = 1.0$ and $h = 12$. When $k = 50$, the sanitized data (under both \emph{Basic} and \emph{Full}) is able to give the exact top 50 most frequent patterns. With the increase of $k$ values, the \emph{accuracy} (the ratio of true positive to $k$) decreases. However, even when $k = 250$, the accuracy of \emph{Full} is still as high as $197/250=78.8\%$. In addition, we can observe that the constrained inference technique again helps obtain more accurate results for frequent sequential pattern mining. When $k=250$, the improvement due to constrained inferences is $10.4\%$.

Table~\ref{table:privacy} presents the utility for frequent sequential pattern mining under different privacy budgets while fixing $h=12$ and $k =200$. Generally, larger privacy budgets lead to more true positives and fewer false positives (false drops). This conforms to the theoretical analysis that a larger privacy budget results in less noise and therefore a more accurate result. Since the most frequent sequential patterns are of small length (typically less than or equal to 3), they have large supports from the underlying database. As a result, the utility is insensitive to varying privacy budgets, and the accuracy is high even when the privacy budget is small. When $\epsilon = 0.5$, the accuracy of \emph{Full} is $160/200=80\%$. Furthermore, it can be learned again that constrained inferences are also helpful for frequent sequential pattern mining.

Table~\ref{table:height} studies how the utility varies under different noisy prefix tree height values with $\epsilon = 1.0$ and $k = 200$. It is interesting to see that under \emph{Basic} the best result is obtained when $h=6$. This again confirms our analysis that truncating the tailing parts of long trajectories has very limited impact on the results of frequent sequential pattern mining. For small $h$ values, due to a larger portion of privacy budget assigned, the construction of each level is more precise, and therefore the accuracy is even higher. With constrained inferences, we are able to obtain better and, more importantly, more stable results. This is vital because under the same $h$ value (e.g., 10-16) the release can simultaneously maintain high utility for both count queries and frequent sequential pattern mining.

\begin{table}
    \caption{Utility for frequent sequential pattern mining vs. $k$ value} \label{table:support}     
    \centering
    \begin{tabular}{|c|c|c|c|c|}\hline
        & True & False Positives & True & False Positives \\ 
       $k$ & Positives & (False Drops) & Positives & (False Drops) \\
        & \emph{Basic} & \emph{Basic}  & \emph{Full} &  \emph{Full} \\ \hline
       50  & 50 & 0 & 50 &  0  \\ \hline
       100  & 92 & 8 & 96 &  4  \\ \hline
       150  &  131 & 19 & 139 &  11  \\ \hline
       200  & 150 & 50 & 169 &  31  \\ \hline
       250  & 171 & 79 & 197 &  53  \\ \hline
    \end{tabular}
\end{table}

\begin{table}
    \caption{Utility for frequent sequential pattern mining vs. privacy budget} \label{table:privacy}     
    \centering
    \begin{tabular}{|c|c|c|c|c|}\hline
       & True & False Positives & True & False Positives \\ 
       $\epsilon$ & Positives & (False Drops) & Positives & (False Drops) \\
        & \emph{Basic} & \emph{Basic} & \emph{Full} &  \emph{Full} \\ \hline
       0.5  & 145 & 55 & 160 &  40  \\ \hline
       0.75  & 147 & 53 & 161 &  39  \\ \hline
       1.0  &  150 & 50 & 169 &  31  \\ \hline
       1.25  & 153 & 47 & 169 &  31  \\ \hline
       1.5  & 153 & 47 & 170 &  30  \\ \hline
    \end{tabular}
\end{table}

\begin{table}
    \caption{Utility for frequent sequential pattern mining vs. height value} \label{table:height}     
    \centering
    \begin{tabular}{|c|c|c|c|c|}\hline
       & True & False Positives & True & False Positives \\ 
       $h$ & Positives & (False Drops) & Positives & (False Drops) \\
        & \emph{Basic} & \emph{Basic}  & \emph{Full} &  \emph{Full} \\ \hline
       6  & 154 & 46 & 167 &  33  \\ \hline
       8  & 154 & 46 & 168 &  32  \\ \hline
       10  &  153 & 47 & 168 &  32  \\ \hline
       12  & 150 & 50 & 169 &  31  \\ \hline
       14  & 150 & 50 & 169 &  31  \\ \hline
       16  & 148 & 52 & 168 &  32  \\ \hline  
       18  & 145 & 55 & 167 &  33  \\ \hline 
       20  & 146 & 54 & 167 &  33  \\ \hline      
    \end{tabular}
\end{table}

\subsection{Scalability}
\begin{figure}[t]
\centering $\begin{array}{c c}
    \multicolumn{1}{l}{\mbox{\bf }} & \multicolumn{1}{l}{\mbox{\bf }} \\
    \hspace{-4mm} \scalebox{0.27}{\includegraphics[bb= 54 202 524 630]{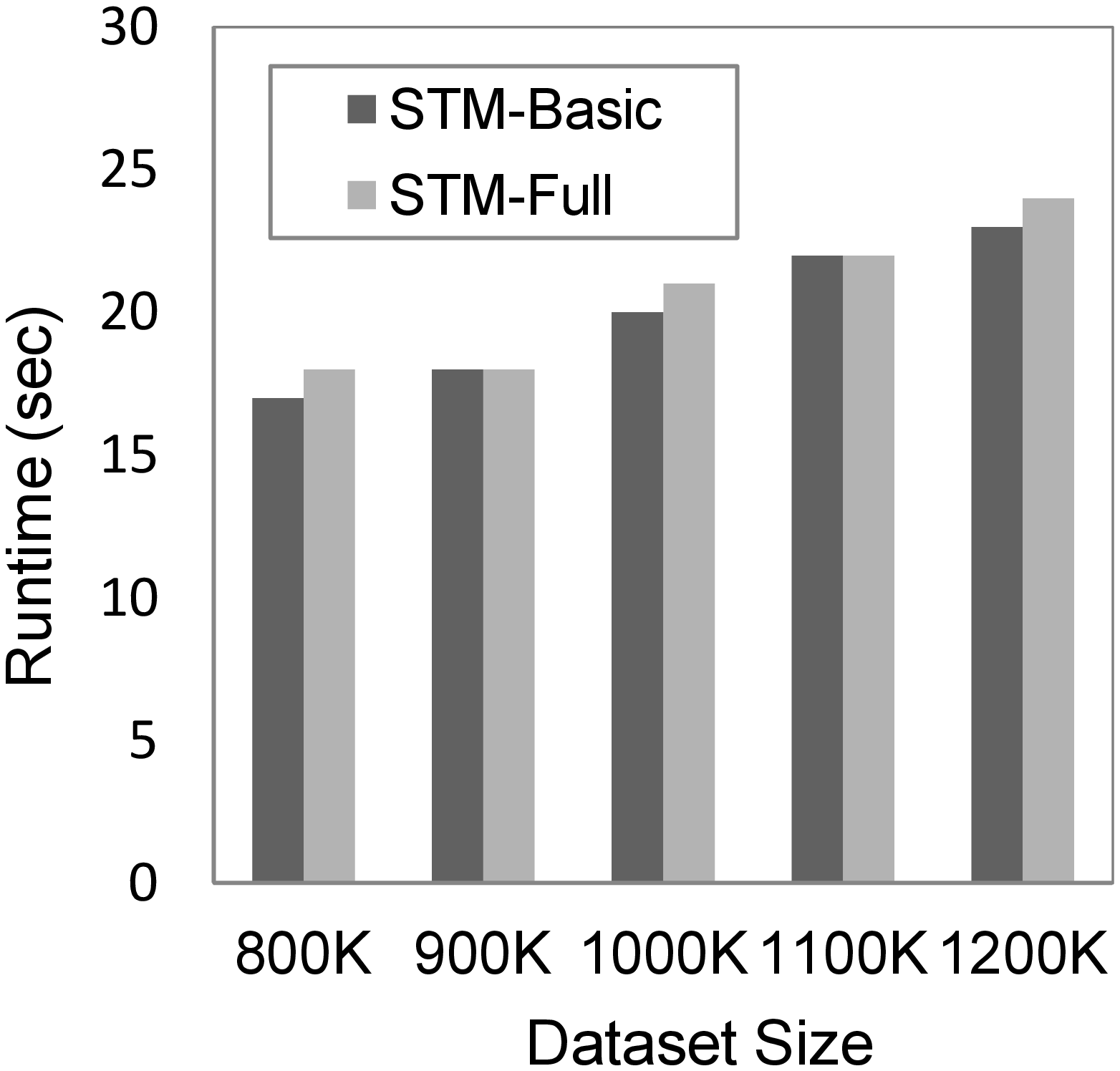}} & \scalebox{0.27}{\includegraphics[bb= 54 202 524 630]{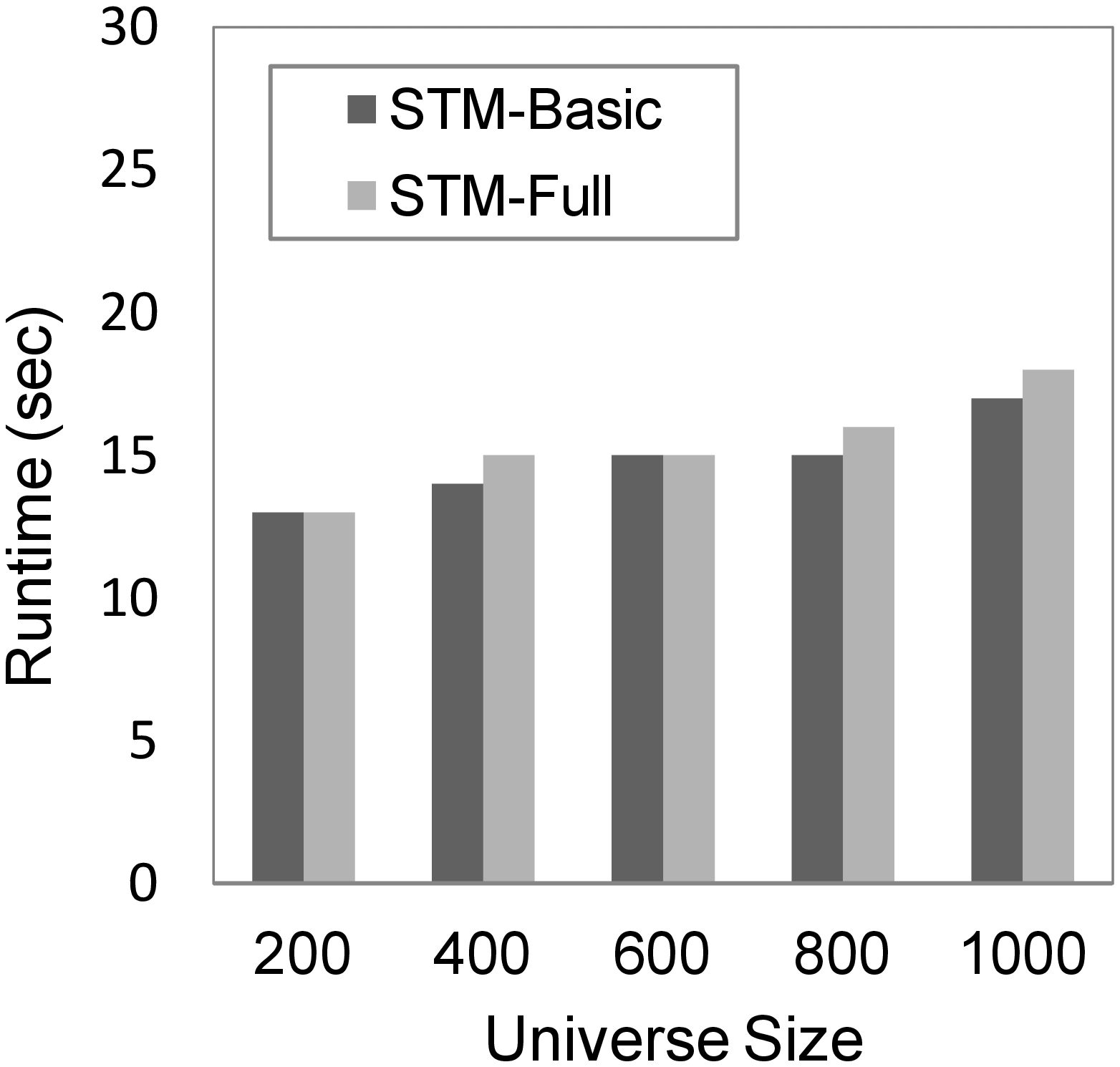}} \\ 
    \mbox{\small{(a) Runtime vs. $|\mathcal{D}|$}} & \mbox{\small{(b) Runtime vs. $|\mathcal{L}|$}}\\
    \end{array}$
\caption{Runtime vs. different parameters.}\label{figure:runtime}
\end{figure}

In the last set of experiments, we examine the scalability of our approach, one of the most important improvement over data-independent approaches. According to the analysis in Section~\ref{subsec:analysis}, the runtime complexity of our approach is dominated by the database size $|\mathcal{D}|$ and the location universe size $|\mathcal{L}|$. Therefore, we study the runtime under different database sizes and different location universe sizes for both \emph{Basic} and \emph{Full}. Figure~\ref{figure:runtime}.a presents the runtime under different database sizes with $\epsilon=1.0$, $h=20$ and $|\mathcal{L}|=1,012$. The test sets are generated by randomly extracting records from the \emph{STM} dataset. We can observe that the runtime is linear to the database size. When $|\mathcal{D}| = 1,200,000$, the runtime of \emph{Full} is just 24 seconds. Moreover, it can be observed that the computational cost of constrained inferences is negligible. This further confirms the benefit of conducting constrained inferences. Figure~\ref{figure:runtime}.b shows how the runtime varies under different location universe sizes. For each universe size, we remove all locations falling out of the universe from the \emph{STM} dataset. This results in a smaller database size. Consequently, we fix the database size for all test sets to 800,000. Again, it can be observed that the runtime scales linearly with the location universe size and that the computational cost of constrained inferences is very small and stable under different location universe sizes. As a summary, our approach is scalable to large trajectory datasets. It takes less 25 seconds to sanitize the \emph{STM} dataset in all previous experiments.

\section{Conclusion} \label{sec:conclusion}

All existing techniques for privacy-preserving trajectory data publishing are derived using partition-based privacy models, which have been shown failing to provide sufficient privacy protection. In this paper, motivated by the STM case, we study the problem of publishing trajectory data in the framework of differential privacy. For the first time, we present a practical data-dependent solution for sanitizing large-scale real-life trajectory data. In addition, we develop a constrained inference technique in order to better the resulting utility. We stress that our approach applies to different types of trajectory data. We believe that our solution could benefit many sectors that face the dilemma between the demands of trajectory data publishing and privacy protection. In future work, we intend to investigate the utility of sanitized data on other data mining tasks, for example, classification and clustering.

\section*{Acknowledgment}

The research is supported in part by the new researchers start-up program from Le Fonds qu\'{e}b\'{e}cois de la recherche sur la nature et les technologies (FQRNT), Discovery Grants (356065-2008) and Canada Graduate Scholarship from the Natural Sciences and Engineering Research Council of Canada (NSERC). We would like to thank the \emph{Soci\'{e}t\'{e} de transport de Montr\'{e}al} for providing real-life trajectory data for experimental evaluation. We would also like to thank Yasuo Tabei for providing the implementation of PrefixSpan.

\bibliographystyle{IEEEtran}

\begin{thebibliography}{10}
\providecommand{\url}[1]{#1}
\csname url@samestyle\endcsname
\providecommand{\newblock}{\relax}
\providecommand{\bibinfo}[2]{#2}
\providecommand{\BIBentrySTDinterwordspacing}{\spaceskip=0pt\relax}
\providecommand{\BIBentryALTinterwordstretchfactor}{4}
\providecommand{\BIBentryALTinterwordspacing}{\spaceskip=\fontdimen2\font plus
\BIBentryALTinterwordstretchfactor\fontdimen3\font minus
  \fontdimen4\font\relax}
\providecommand{\BIBforeignlanguage}[2]{{%
\expandafter\ifx\csname l@#1\endcsname\relax
\typeout{** WARNING: IEEEtran.bst: No hyphenation pattern has been}%
\typeout{** loaded for the language `#1'. Using the pattern for}%
\typeout{** the default language instead.}%
\else
\language=\csname l@#1\endcsname
\fi
#2}}
\providecommand{\BIBdecl}{\relax}
\BIBdecl

\bibitem{UAW06}
M.~Utsunomiya, J.~Attanucci, and N.~Wilson, ``Potential uses of transit smart
  card registration and transaction data to improve transit planning,''
  \emph{Transportation Research Record: Journal of the Transportation Research
  Board}, no. 1971, pp. 119--126, 2006.

\bibitem{PK08}
J.~Y. Park and D.~J. Kim, ``The potential of using smart card data to define
  the use of public transit in seoul,'' \emph{Transportation Research Record:
  Journal of the Transportation Research Board}, no. 2063, pp. 3--9, 2008.

\bibitem{BCM06}
A.~Bruno, M.~Catherine, and T.~Martin, ``Mining public transport user behaviour
  from smart card data,'' in \emph{Proceedings of 12th IFAC Symposium on
  Information Control Problems in Manufacturing}, St-\'Etienne, France, May
  2006.

\bibitem{MMC09}
P.~Marie-Pierre, T.~M., and M.~Catherine, ``Smart card data use in public
  transit: A literature review,'' \emph{Transportation Research C: Emerging
  Technologies}, vol.~19, no.~4, pp. 557--568, 2011.

\bibitem{CMB07}
M.~Catherine, T.~Martin, and A.~Bruno, ``Measuring transit performance using
  smart card data,'' in \emph{Proceedings of the 11th World Conference on
  Transportation Research}, Berkeley, USA, Jun. 2007.

\bibitem{RC01}
R.~Clarke, ``Person location and person tracking: Technologies, risks, and
  policy implications,'' \emph{Information Technology \& People}, vol.~14,
  no.~2, pp. 206--231, 2001.

\bibitem{ABN08}
O.~Abul, F.~Bonchi, and M.~Nanni, ``Never walk alone: Uncertainty for anonymity
  in moving objects databases,'' in \emph{Proceedings of the 24th IEEE
  International Conference on Data Engineering}, Cancun, Mexico, Apr. 2008, pp.
  376--385.

\bibitem{TM08}
M.~Terrovitis and N.~Mamoulis, ``Privacy preservation in the publication of
  trajectories,'' in \emph{Proceedings of the 9th International Conference on
  Mobile Data Management}, Beijing, China, Apr. 2008, pp. 65--72.

\bibitem{YBLW09}
R.~Yarovoy, F.~Bonchi, L.~V.~S. Lakshmanan, and W.~H. Wang, ``Anonymizing
  moving objects: How to hide a {MOB} in a crowd?'' in \emph{Proceedings of the
  12th International Conference on Extending Database Technology},
  Saint-Petersburg, Russia, Mar. 2009, pp. 72--83.

\bibitem{HXODN10}
H.~Hu, J.~Xu, S.~T. On, J.~Du, and J.~K.-Y. Ng, ``Privacy-aware location data
  publishing,'' \emph{ACM Transactions on Database Systems}, vol.~35, no.~3,
  p.~17, 2010.

\bibitem{CFMD11}
R.~Chen, B.~C.~M. Fung, N.~Mohammed, and B.~C. Desai, ``Privacy-preserving
  trajectory data publishing by local suppression,'' \emph{Information
  Sciences}, under minor modification.

\bibitem{MA10}
A.~Monreale, G.~Andrienko, N.~Andrienko, F.~Giannotti, D.~Pedreschi,
  S.~Rinzivillo, and S.~Wrobel, ``Movement data anonymity through
  generalization,'' \emph{Transactions on Data Privacy}, vol.~3, no.~2, pp.
  91--121, 2010.

\bibitem{GKS08}
S.~R. Ganta, S.~P. Kasiviswanathan, and A.~Smith, ``Composition attacks and
  auxiliary information in data privacy,'' in \emph{Proceedings of the 14th ACM
  SIGKDD International Conference on Knowledge Discovery and Data Mining}, Las
  Vegas, USA, Aug. 2008, pp. 265--273.

\bibitem{Sweeney02}
L.~Sweeney, ``{$k$}-anonymity: A model for protecting privacy,''
  \emph{International Journal on Uncertainty, Fuzziness and Knowledge-based
  Systems}, vol.~10, no.~5, pp. 557--570, 2002.

\bibitem{Kifer09}
D.~Kifer, ``Attacks on privacy and definetti's theorem,'' in \emph{Proceedings
  of the ACM SIGMOD International Conference on Management of Data},
  Providence, USA, Jun. 2009, pp. 127--138.

\bibitem{WFWYP}
R.~C.~W. Wong, A.~Fu, K.~Wang, P.~S. Yu, and J.~Pei, ``Can the utility of
  anonymized data be used for privacy breaches,'' \emph{ACM Transactions on
  Knowledge Discovery from Data}, to appear.

\bibitem{DMNS06}
C.~Dwork, F.~McSherry, K.~Nissim, and A.~Smith, ``Calibrating noise to
  sensitivity in private data analysis,'' in \emph{Proceedings of 3rd Theory of
  Cryptography Conference}, New York, USA, Mar. 2006, pp. 265--284.

\bibitem{KM11}
D.~Kifer and A.~Machanavajjhala, ``No free lunch in data privacy,'' in
  \emph{Proceedings of the ACM SIGMOD International Conference on Management of
  Data}, Athens, Greece, Jun. 2011.

\bibitem{BLR08}
A.~Blum, K.~Ligett, and A.~Roth, ``A learning theory approach to
  non-interactive database privacy,'' in \emph{Proceedings of the 40th Annual
  ACM Symposium on Theory of Computing}, Victoria, Canada, May 2008, pp.
  609--618.

\bibitem{DNRRV09}
C.~Dwork, M.~Naor, O.~Reingold, G.~N. Rothblum, and S.~Vadhan, ``On the
  complexity of differentially private data release: Efficient algorithms and
  hardness results,'' in \emph{Proceedings of the 41st Annual ACM Symposium on
  Theory of Computing}, Bethesda, USA, May 2009, pp. 381--390.

\bibitem{XWG10}
X.~Xiao, G.~Wang, and J.~Gehrke, ``Differential privacy via wavelet
  transforms,'' in \emph{Proceedings of the 26th International Conference on
  Data Engineering}, Long Beach, USA, Mar. 2010, pp. 225--236.

\bibitem{XXY10}
Y.~Xiao, L.~Xiong, and C.~Yuan, ``Differentially private data release through
  multidimensional partitioning,'' in \emph{Proceedings of the Secure Data
  Management, 7th VLDB workshop}, Singapore, Sep. 2010, pp. 150--168.

\bibitem{MCFY11}
N.~Mohammed, R.~Chen, B.~C.~M. Fung, and P.~S. Yu, ``Differentially private
  data release for data mining,'' in \emph{Proceedings of the 17th ACM SIGKDD
  International Conference on Knowledge Discovery and Data Mining}, San Diego,
  USA, Aug. 2011.

\bibitem{CMFDX11}
R.~Chen, N.~Mohammed, B.~C.~M. Fung, B.~C. Desai, and L.~Xiong, ``Publishing
  set-valued data via differential privacy,'' \emph{Proceedings of the VLDB
  Endowment}, vol.~4, no.~11, 2011.

\bibitem{AS95}
R.~Agrawal and R.~Srikant, ``Mining sequential patterns,'' in \emph{Proceedings
  of the 11th IEEE International Conference on Data Engineering}, Taipei,
  Taiwan, Mar. 1995, pp. 3--14.

\bibitem{HRMS10}
M.~Hay, V.~Rastogi, G.~Miklau, and D.~Suciu, ``Boosting the accuracy of
  differentially private histograms through consistency,'' \emph{Proceedings of
  the VLDB Endowment}, vol.~3, no.~1, pp. 1021--1032, 2010.

\bibitem{LHRMM10}
C.~Li, M.~Hay, V.~Rastogi, G.~Miklau, and A.~McGregor, ``Optimizing linear
  counting queries under differential privacy,'' in \emph{Proceedings of the
  29th ACM SIGMOD-SIGACT-SIGART Symposium on Principles of Database Systems},
  Indianapolis, USA, Jun. 2010, pp. 123--134.

\bibitem{RR10}
A.~Roth and T.~Roughgarden, ``Interactive privacy via the median mechanism,''
  in \emph{Proceedings of the 42nd ACM Symposium on Theory of Computing},
  Cambridge, USA, Jun. 2010, pp. 765--774.

\bibitem{XBHG11}
X.~Xiao, G.~Bender, M.~Hay, and J.~Gehrke, ``ireduct: Differential privacy with
  reduced relative errors,'' in \emph{Proceedings of the ACM SIGMOD
  International Conference on Management of Data}, Athens, Greece, Jun. 2011.

\bibitem{BCDKMT07}
B.~Barak, K.~Chaudhuri, C.~Dwork, S.~Kale, F.~McSherry, and K.~Talwar,
  ``Privacy, accuracy, and consistency too: A holistic solution to contingency
  table release,'' in \emph{Proceedings of the 26th ACM SIGACT-SIGMOD-SIGART
  Symposium on Principles of Database Systems}, Beijing, China, Jun. 2007, pp.
  273--282.

\bibitem{BLST2010}
R.~Bhaskar, S.~Laxman, A.~Smith, and A.~Thakurta, ``Discovering frequent
  patterns in sensitive data,'' in \emph{Proceedings of the 16th ACM SIGKDD
  International Conference on Knowledge Discovery and Data Mining}, Washington,
  USA, Jul. 2010, pp. 503--512.

\bibitem{FS10}
A.~Friedman and A.~Schuster, ``Data ming with differential privacy,'' in
  \emph{Proceedings of the 16th ACM SIGKDD International Conference on
  Knowledge Discovery and Data Mining}, Washington, USA, Jul. 2010, pp.
  493--502.

\bibitem{KKMN09}
A.~Korolova, K.~Kenthapadi, N.~Mishra, and A.~Ntoulas, ``Releasing search
  queries and clicks privately,'' in \emph{Proceedings of the 18th
  International Conference on World Wide Web}, Madrid, Spain, Apr. 2009, pp.
  171--180.

\bibitem{MKAGV08}
A.~Machanavajjhala, D.~Kifer, J.~M. Abowd, J.~Gehrke, and L.~Vilhuber,
  ``Privacy: Theory meets practice on the map,'' in \emph{Proceedings of the
  24th IEEE International Conference on Data Engineering}, Cancun, Mexico, Apr.
  2008, pp. 277--286.

\bibitem{Dwork2011}
C.~Dwork, ``A firm foundation for private data analysis,'' \emph{Communications
  of the ACM}, vol.~54, no.~1, pp. 86--95, 2011.

\bibitem{McSherry09}
F.~McSherry, ``Privacy integrated queries: An extensible platform for
  privacy-preserving data analysis,'' in \emph{Proceedings of the ACM SIGMOD
  International Conference on Management of Data}, Providence, USA, Jun. 2009,
  pp. 19--30.

\bibitem{MT07}
F.~McSherry and K.~Talwar, ``Mechanism design via differential privacy,'' in
  \emph{Proceedings of the 48th Annual IEEE Symposium on Foundations of
  Computer Science}, Providence, USA, Oct. 2007, pp. 94--103.

\bibitem{ESAG04}
A.~V. Evfinievski, R.~Srikant, R.~Agrawal, and J.~Gehrke, ``Privacy preserving
  mining of association rules,'' \emph{Information Systems}, vol.~29, no.~4,
  pp. 343--364, 2004.

\bibitem{CPS11}
G.~Cormode, M.~Procopiuc, and D.~Srivastava, ``Differentially private
  publication of spare data,'' in \emph{CoRR}, Mar. 2011.

\bibitem{GRS09}
A.~Ghosh, T.~Roughgarden, and M.~Sundararajan, ``Universally utility-maximizing
  privacy mechanisms,'' in \emph{Proceedings of the 41st Annual ACM Symposium
  on Theory of Computing}, Bethesda, USA, May 2009, pp. 3--14.

\bibitem{JRT97}
J.~R. Taylor, \emph{An Introduction to Error Analysis}.\hskip 1em plus 0.5em
  minus 0.4em\relax University Science Books, 1997.

\bibitem{PHMPCDH01}
J.~Pei, J.~Han, B.~Mortazavi-Asl, H.~Pinto, Q.~Chen, U.~Dayal, and M.~Hsu:,
  ``Prefixspan: Mining sequential patterns by prefix-projected growth,'' in
  \emph{Proceedings of the 17th International Conference on Data Engineering},
  Heidelberg, Germany, Apr. 2001, pp. 215--224.

\end{thebibliography}

\end{document}